\documentclass[aip,jcp,reprint]{revtex4-2}
\usepackage[T1]{fontenc} 
\usepackage[english]{babel}
\usepackage[nolist]{acronym}
\usepackage{amsmath,amsfonts,amssymb}
\usepackage{bm}
\usepackage{dcolumn}
\usepackage{graphicx}
\usepackage[colorlinks=true, allcolors=blue]{hyperref}
\usepackage{physics}
\usepackage{stmaryrd}

\renewcommand{\i}{\mathrm{i}}
\newcommand{\tj}[6]{ 
	\begin{pmatrix}
		#1 & #2 & #3 \\
		#4 & #5 & #6 
	\end{pmatrix}}
\newcommand{\sj}[6]{ 
	\begin{Bmatrix}
		#1 & #2 & #3 \\
		#4 & #5 & #6 
	\end{Bmatrix}}
\newcommand{\itens}[3]{\hat{#1}^{#2}_{#3}}
\newcommand{\redmel}[3]{\langle #1 || #2 || #3 \rangle}
\newcommand{\smmat}[1]{\langle\pmb{\rho}^{#1}\rangle}
\renewcommand{\vec}[1]{\mathbf{#1}}
\newcommand{\Vmat}{\pmb{V}}
\newcommand{\Vop}{\hat V}
\newcommand{\smel}[2]{\langle\rho^{#1}_{#2}\rangle}
\newcommand{\Ti}{TiCl$_4$}
\newcommand{\FeWat}{[Fe(H$_2$O)$_6$]$^{2+}$}
\newcommand{\CrWat}{[Cr(H$_2$O)$_6$]$^{3+}$}

\newcommand{\Ssq}{$\langle \hat S^2 \rangle$}
\newcommand{\Ledge}{L$_{2,3}$}

\begin{document}

\title{Spin-flip dynamics in core-excited states in the basis of irreducible spherical tensor operators}

\author{Thies Romig}
\affiliation{Institut f\"{u}r Physik, Universit\"{a}t Rostock, A.-Einstein-Strasse 23-24, 18059 Rostock, Germany}

\author{Vladislav Kochetov}
\affiliation{Institut f\"{u}r Physik, Universit\"{a}t Rostock, A.-Einstein-Strasse 23-24, 18059 Rostock, Germany}

\author{Sergey I. Bokarev}
\email{sergey.bokarev@tum.de}
\affiliation{Chemistry Department, School of Natural Sciences, Technical University of Munich, Lichtenbergstr. 4, 85748 Garching, Germany}
\affiliation{Institut f\"{u}r Physik, Universit\"{a}t Rostock, A.-Einstein-Strasse 23-24, 18059 Rostock, Germany}

\date{\today}

\begin{abstract}
    Recent experimental advances in ultrafast science put different processes occurring on the electronic timescale below a few femtoseconds in focus.
    In the present theoretical work, we demonstrate how the transformation and propagation of the density matrix in the basis of irreducible spherical tensors can be conveniently used to study sub-few fs spin-flip dynamics in the core-excited transition metal compounds.
    With the help of the Wigner-Eckart theorem, such a transformation separates the essential dynamical information from the geometric factors governed by the angular momentum algebra. 
    We show that an additional reduction can be performed by the physically motivated truncation of the spherical tensor basis.
    In particular, depending on the degree of coherence, the ultrafast dynamics can be considered semi-quantitative in the notably reduced spherical basis when only total populations of the basis states of the given spin are of interest.
    Such truncation should be especially beneficial when the number of the high-spin basis states is vast, as it substantially reduces computational costs.
\end{abstract}

\maketitle

\begin{acronym}
	\acro{AMFI}{Atomic Mean Field Integrals}
	\acro{AS}{Active Space}
	\acro{CAS}{Complete Active Space}
	\acro{CI}{Configuration Interaction}
	\acro{CSF}{Configuration State Function}
	\acro{EOM}{Equation of Motion}
	\acro{HF}{Hartree-Fock}
	\acro{HHG}{High Harmonics Generation}
	\acro{LvN}{Liouville-von-Neumann}
	\acro{MCSCF}{Multi-configurational Self-Consistent Field}
	\acro{MO}{Molecular Orbital}
	\acro{RAS}{Restricted Active Space}
	\acro{RASPT2}{Restricted Active Space Second Order Perturbation Theory}
	\acro{RASSCF}{Restricted Active Space Self-Consistent Field}
	\acro{RASSI}{Restricted Active Space State Interaction}
	\acro{RIXS}{Resonant Inelastic X-ray Scattering}
	\acro{SCF}{Self-Consistent Field}
	\acro{SF}{Spin--Free}
	\acro{SO}{Spin--Orbit}
	\acro{SOC}{Spin--Orbit Coupling}
	\acro{TD-RASCI}{Time-Dependent Restricted Active Space Configuration Interaction}
	\acro{WE}{Wigner--Eckart}
	\acro{XAS}{X-ray Absorption Spectrum}
	\acro{XES}{X-ray Emission Spectrum}
	\acro{XFEL}{X-ray Free Electron Laser}
\end{acronym}

\section{Introduction}
\label{sec:intro}

Electron dynamics is a fundamental phenomenon determining the mechanism of many processes in molecules and condenced matter.~\cite{Halcrow2013,Schultz2014,Young_JPB_2018,Bande_CM_2022}
It represents an initial elementary step involving electronic reorganization, while nuclei start to move only in response to it, leading, in turn, to electronic decoherence.
One of the most important effects is charge transfer,~\cite{May2011,Worner_SD_2017} playing a decisive role in many chemical and biochemical transformations, such as photosynthesis, with the electron correlation- and relaxation-driven charge migration~\cite{Kuleff_JPB_2014,Nisoli_CR_2017} being its elementary step.
Electronic response to strong fields can also lead to highly non-linear processes such as high harmonic generation.~\cite{Hentschel_N_2001,Corkum_NP_2007}

Another process that is driven by electronic coupling is spin-flip dynamics.~\cite{Marian_WCMS_2012,Mai_CS_2019a,Jiang_NC_2020}
In the weak interaction regime, it should be necessarily driven by nuclear motion to bring interacting states close enough for the intersystem crossing to happen.
The situation changes when the coupling is strong due to electronic excitation from core shells with non-zero angular momentum, e.g., 2p levels.~\cite{Kochetov_JCP_2020, Kochetov_JCTC_2022}
In this case, the coupling may be larger, starting from several eV for lighter elements, and does not necessarily require a nuclear motion for spin-flip to happen.~\cite{Wang_PRL_2017}
The spin dynamics can be especially intricate when successive ionizations occur under ultra-intense X-ray irradiation.~\cite{Rudenko_N_2017}
In this case, one has to consider multiple spin manifolds calling for an efficient theoretical treatment.

One of the ways is to separate geometric, due to the angular symmetry of the system, and essential dynamical parts of the time-evolution of the system's wave function or density matrix.
It can be afforded, for instance, by expanding the density operator in terms of irreducible tensor operators.~\cite{Blum2012}
Fano first suggested the systematic use of tensor operators,~\cite{Fano_PR_1953} and later their use was extensively enlarged to applications in angular correlation theory in atomic physics~\cite{Steffen1975,Blum_PR_1979} and the description of atom-light interactions.~\cite{Fano_RMP_1973,Omont_PQE_1977,Hertel_AiAaMP_1978,Givon_PRL_2013}

When talking about a large number of states of different multiplicities, one might be interested in some reduced representation of a system that should factor geometric parts out.
In an energy (state) basis, the zero-order pure-spin states are characterized by their spin $S$ and its projection onto the quantization axis $M$.
When several thousands of states are in focus, i.e., when the state density and the width of the excitation pulse are significant, one would wish to neglect the dynamics of magnetic sublevels and is primarily interested in the population of the groups of these sublevels belonging to the same spin-free state or even collectively the population of all the states with the same spin.
The averaging over $M$ reduces the information available and should require simplified propagation, which can be still very costly for many states.

Here, we attempt such a description using the spherical tensor basis to represent the density matrix employing the density-matrix-based \ac{TD-RASCI} method.~\cite{Wang_MP_2017, Kochetov_JCTC_2022}
In particular, we consider the ultrafast dynamics in three transition metal compounds, \Ti, \FeWat, and \CrWat.
Upon an excitation from the ground state of these systems to a superposition of \Ledge\ 2p-hole core-excited states, the ultrafast spin-mixing occurs, resulting in a substantial population of spin-flipped states of different multiplicities.
Finally, we note that the method is implemented in the \texttt{OpenMolcas} program package~\cite{LiManni_JCTC_2023} in the module \texttt{RhoDyn},~\cite{Kochetov_JCTC_2022} allowing for both full and reduced propagation in the state or spherical tensor bases.

\section{Theory and implementation}
\label{sec:theory}

\subsection{Equation of Motion (EOM) in the tensor basis}
\label{subsec:eom}

Below we utilize the density-matrix-based \ac{TD-RASCI} formalism.~\cite{Kochetov_JCTC_2022}
The state of the system is characterized by a density operator $\hat \rho$, the time evolution of which obeys the \ac{LvN} equation of motion
\begin{equation}\label{eq:LvN}
	\dot{\hat\rho} = -\i [\hat H_{\rm CI} + \Vop - \hat{\pmb{\mu}}\cdot\pmb{\mathcal{E}}, \hat\rho]\,.
\end{equation}
Here, the Hamiltonian is split into three terms: $\hat H_\text{CI}$ represents the interelectronic interaction and thus governs dynamics driven by electronic correlation and relaxation. 
This part is treated here through the \ac{CI} method.
Next, $\Vop$ is the \ac{SOC} operator, and the last semi-classical term corresponds to the interaction of the system with an external time-dependent electric field $\pmb{\mathcal{E}}$ via its dipole moment $\hat{\pmb{\mu}}$, i.e., employing long-wavelength approximation.
In this article, nuclei stay fixed, and we neglect energy and phase relaxation processes for simplicity of discussion, i.e., the system stays closed at all times; this limitation can be readily lifted,~\cite{Kochetov_JCP_2020} thanks to the density-matrix formulation of the theory.

The original pure-spin state basis consists of the states $\ket{aSM}$, where $a$ is the index of the spatial part collectively denoting all relevant quantum numbers, and $S$ and $M$ are the total spin and its projection onto the quantization axis.
One can consider the spatial part as being, e.g., the eigenstates of $\hat H_\text{CI}$, which will be called \ac{SF} states below, while the $\ket{aSM}$ states will be called basis spin states.
The \ac{SF} states have no explicit $M$-dependence and thus need to be extended to include the respective spin part when spin-dependent interactions are to be considered.
It increases the state vector size by the spin multiplicity factor and the density matrix by its square.
We separate spatial and spin parts in the basis states $\ket{aS}\otimes\ket{SM}$;
the total spin naturally relates to the spin part but is fixed by the spatial index $a$, so the summation over $a$ will be in the following denoted as a summation over $aS$.
Thus, the spin-independent operators attain the structure $\hat H_{\rm CI}\otimes\hat 1$ and $\hat{\pmb{\mu}}\cdot\pmb{\mathcal{E}} \otimes \hat 1$. In contrast, $\Vop$ couples both parts and can be represented as
\begin{equation}\label{eq:Vop}
	\Vop=\sum_{m=0,\pm 1} (-1)^m \hat L^1_{-m} \otimes \hat S^1_m\,,
\end{equation}
where $L^1_{(0,\pm 1)}$ and $S^1_{(0,\pm 1)}$ are standard components of angular momentum and spin tensor operators of rank 1.

Assuming this setting, the uncorrelated initial density operator can be represented as the direct product of spatial $\hat\pi(0)$ and spin $\hat{\sigma}(0)$ parts at time $t=0$ 
\begin{equation}
	\hat\rho(0)=\hat\pi(0)\otimes\hat{\sigma}(0)\,.
\end{equation}
At later times when \ac{SOC} is switched on, the factorization does not hold.
However, when writing these operators in the matrix form in a state basis, one can consider this structure to still hold within the subblocks of the total density matrix.
Further, we will consider only a separate block 
$\pmb{\rho}_{aS,bS'}=\pi_{ab}\otimes\pmb{\sigma}_{SS'}$ and express the $\pmb{\sigma}$ matrix in terms of irreducible tensor matrix elements $\vec{T}^{k}_{q}(S,S')$.
Note that $\pmb{\rho}_{aS,bS'}$ and $\pmb{\sigma}_{SS'}$ denote the whole $(2S+1)\times(2S'+1)$ subblocks and $\pi_{a,b}$ a single spatial matrix element.
With this, the entire block can be written as
\begin{equation}\label{eq:tens_exp}
	\pmb{\rho}_{aS,bS'}=\sum_{kq}\langle\rho^{k\,q}_{aS,bS'}\rangle \vec{T}^{k}_{q}(S,S')
\end{equation}
Here, the dynamical factor (state multipole) $\langle\rho^{k\,q}_{aS,bS'}\rangle=\pi_{ab}\cdot\langle\sigma^{k\,q}(S,S')\rangle$ absorbs both the spatial part $\pi_{ab}$ and the expansion coefficient $\langle{\sigma}^{k\,q}(S,S')\rangle=\Tr{\hat\sigma \hat{T}^{k\dagger}_{q}(S,S')}$, where the trace is taken with respect to the $\ket{SM}$ spin basis states.
To do so, we recall that for the matrix element of the irreducible spherical tensor operator 
$\hat{T}^{k}_{q}$ of rank $k$ and projection $q$, the relation
\begin{multline}\label{eq:t_kq}
	\bra{SM}\hat T^{k}_{q}(S,S')\ket{S'M'} = \\ (-1)^{S-M}\sqrt{2k+1}\tj{S}{k}{S'}{-M}{q}{M'}\,,
\end{multline}
holds according to the \ac{WE} theorem;~\cite{Edmonds1957} parentheses denote the Wigner 3j symbol.

Next, we notice that 
\begin{equation}\label{eq:H_CI}
	\bra{aSM}\hat H_{\rm CI} \otimes \hat 1 \ket{bS'M'}=E^{\rm SF}_a\delta_{ab}\delta_{SS'}\delta_{MM'}
\end{equation}
\begin{equation}\label{eq:muE}
	\bra{aSM}\vec{\hat{\pmb{\mu}}}\cdot\pmb{\mathcal{E}} \otimes \hat 1 \ket{bS'M'}=\pmb{\mu}^{\rm SF}_{ab}\cdot\pmb{\mathcal{E}}\ \delta_{SS'}\delta_{MM'}\,;
\end{equation}
since energies and transition dipole matrix elements are estimated solely in the \ac{SF} basis, they attain the respective label.
For \ac{SOC}, we write
\begin{align}\label{eq:V_SOC}
	&\bra{aSM}\Vop\ket{bS'M'} = \nonumber\\
	&\sum_{m=0,\pm1}(-1)^m \bra{aS}\itens{L}{1}{-m}\ket{bS'}\bra{SM}\itens{S}{1}{m}\ket{S'M'} = \nonumber \\
	&\sqrt{3}\sum_{m=0,\pm1}(-1)^{S-M+m} \tj{S}{1}{S'}{-M}{m}{M'}\redmel{aS}{ \Vop^m}{bS'} \, ,
\end{align}
where 
\begin{equation}\label{eq:SOC_red}
\redmel{aS}{\Vop^m}{bS'}=V^m_{aS,bS'}=\bra{a}\itens{L}{1}{-m}\ket{b}\redmel{S}{\itens{S}{1}{}}{S'}
\end{equation}
is the \ac{WE} semi-reduced \ac{SOC} matrix element; note the residual dependence on the $m$ number.
These elements form a reduced \ac{SOC} matrix (tensor) denoted below as $\mathbb{V}=(\Vmat^m)$.

We obtain an equation for the evolution of state multipoles $\langle\rho^{k\,q}_{aS,bS'}\rangle$ by multiplying both sides of the \ac{LvN} Eq.~\eqref{eq:LvN} with the $\hat1\otimes\itens{T}{k^\dagger}{q}(S,S')$ operator, tracing out spin degrees of freedom, and sandwiching between the \ac{SF} $\ket{aS}$ basis states. 
For details of derivation, see Appendix~\ref{sec:appendix}. 

For a general density, the equation for its state multipoles in matrix form reads
\begin{equation}\label{eq:LvN_spherical}
 		\i\langle\dot{\pmb\rho}^{k\,q}\rangle = \left[
		\mathbf H_{\rm CI}
		- \hat{\pmb\mu}
		\cdot\pmb{\mathcal{E}},\smmat{k\,q} \right]
		+ \sum_{\substack{K=k,k\pm1\\|Q|\le K}} \left\{\mathbb{V}, \smmat{KQ}\right\}\, ,   
\end{equation}
where $k=0,\ldots,2\cdot\max\{S,S',S''\},\, q=-k,\ldots,k$, and $K\le 2\cdot\max\{S,S',S''\}$.
The first conventional commutator on the right-hand side of the Eq.~\eqref{eq:LvN_spherical} corresponds to the influence of electron correlation and coupling to the external field.
In the basis of eigenstates of $\hat H_{\rm CI}$, the $\mathbf H_{\rm CI}$ matrix is diagonal with \ac{SF} energies $E^{\rm SF}_a$ on it, Eq.~\eqref{eq:H_CI}.
The commutator-like \ac{SOC} terms in curly braces on the right-hand side of Eq.~\eqref{eq:LvN_spherical} mix state multipoles of different ranks and projections.
These terms  involve weighted matrix-matrix multiplication
\begin{widetext}
	\begin{align}
\left\{\mathbb{V}, \smmat{KQ}\right\}_{aS,bS'} =
\sum_{cS''}\sum_{m=0,\pm1}\mathcal{Y}_1 \, {V}^m_{aS,cS''} \, \langle\rho^{K\,Q}_{cS'',bS'}\rangle +
		\mathcal{Y}_2 \, \langle\rho^{K\,Q}_{aS,cS''}\rangle V^m_{cS'',bS'} \, ,
		\end{align}
with geometric weighting factors depending on all spins, ranks, and their projections via Wigner 3j and 6j symbols
		\begin{align}
		\mathcal{Y}_1&=(-1)^{S+S'-Q} \sqrt{3(2k+1)(2K+1)}\tj{K}{1}{k}{Q}{m}{-q}\sj{K}{1}{k}{S}{S'}{S''}\label{eq:ypsilon1}\\
		\mathcal{Y}_2&=(-1)^{S+S'-Q+K+k}\sqrt{3(2k+1)(2K+1)}\tj{K}{1}{k}{Q}{m}{-q}\sj{1}{K}{k}{S}{S'}{S''}\,.\label{eq:ypsilon2}
	\end{align}
\end{widetext}

In the resulting Eq.~\eqref{eq:LvN_spherical}, the $\smmat{k\,q}$ matrices have dimensions $N_{\rm SF}\times N_{\rm SF}$, where $N_{\rm SF}$ is the number of \ac{SF} states and, thus, the complexity of the problem is reduced approximately by a factor of the squared mean multiplicity.
However, the dependence on $M$ is replaced by the dependence on $k$ and $q$, leading to an increased number of smaller problems.

This \ac{EOM} in terms of the state multipoles is equivalent to the propagation in the basis of $\ket{aSM}$ spin states if one takes all necessary ranks $k$ into account.
Because of this equivalence, one can resolve the dynamics of individual $M$ components of the density matrix delivered by the back transformation from the spherical tensor basis to the spin-state basis.
It allows explicitly treating cases with circularly polarized light, interaction with magnetic fields, or when the initial condition implies significantly different populations of $M$-sublevels.
However, it does improve computational effort, and at least for small and medium numbers of states, the scaling is worse than the propagation in the state basis.
This calls for simplifications and reduction of complexity.

\subsection{Reduced EOM}
\label{subsec:eom_reduced}

If the distribution between different magnetic quantum numbers $M$ is not of interest, only the reduced density matrix $\tilde{\pmb{\rho}}$ must be propagated.
It is obtained from the full density matrix tracing out $M$-dependence $\tilde{\pmb{\rho}}=\Tr\pmb{\rho}$.
The derivation procedure is entirely analogous to that for the full density matrix as sketched in Appendix~\ref{sec:appendix}, but the elements of the tensor operator matrices, Eq.~\eqref{eq:t_kq}, need to be considered explicitly. 
This is owing to an additional condition:
in Eqs.~\eqref{eq:tens_exp} and~\eqref{eq:t_kq}, only summed diagonal elements of $\vec{T}^{k}_{q}(S,S')$ are regarded, i.e., $M=M'$, leading to the expression for the reduced matrix element
\begin{align}
    &\tilde{\rho}_{aS,bS'}=\Tr \pmb\rho_{aS,bS'} \nonumber\\
	&=\sum_M\sum_{k}\langle\rho^{k\,0}_{aS,bS'}\rangle (-1)^{S-M}\sqrt{2k+1}\tj{S}{k}{S'}{-M}{0}{M}\,,\label{eq:reduction}
\end{align}
where the selection rules of the $3j$ symbol demand $q=0$. 
The fact that only $q=0$ contributes for every $k$ is in accord with the interpretation of $\smel{k\,0}{aS,bS'}$ as representing diagonal elements of the original density matrix in the state basis, see Appendix~\ref{app:meaning}.
Taking this into account leads to the reduced \ac{EOM}
\begin{equation}\label{eq:EOM_reduced}
 		\i\langle\dot{\pmb\rho}^{k\,0}\rangle = \left[
		\mathbf H_{\rm CI}
		- \hat{\pmb\mu}
		\cdot\pmb{\mathcal{E}},\smmat{k\,0} \right]
		+ \sum_{\substack{K=k,k\pm1\\|Q|\le 1}} \interleave\mathbb{V}, \smmat{KQ}\interleave   \, ,
\end{equation}
with the slightly modified \ac{SOC} commutator-like term and different geometric factors
\begin{widetext}\begin{align}
    \interleave\mathbb{V}, \smmat{K\,Q}\interleave_{aS, bS'} = \sum_{cS''}\mathcal{X}_1
	V^{-Q}_{aS,cS''}\langle\rho^{K\,Q}_{cS'',bS'}\rangle +\mathcal{X}_2\langle\rho^{K\,Q}_{aS,cS''}\rangle V^{-Q}_{cS'',bS'}\,,
	\end{align}
\begin{align}
    \mathcal{X}_1&=(-1)^{S+S'-Q} \sqrt{3(2k+1)(2K+1)}\tj{K}{1}{k}{Q}{-Q}{0}\sj{K}{1}{k}{S}{S'}{S''}\\
    \mathcal{X}_2&=(-1)^{S+S'-Q+K+k}\sqrt{3(2k+1)(2K+1)}\tj{K}{1}{k}{Q}{-Q}{0}\sj{1}{K}{k}{S}{S'}{S''}\,.
\end{align}
\end{widetext}

However, as the derivative of the $\smmat{k\,0}$ also depends on matrices with $Q=0,\pm1$, Eq.~\eqref{eq:EOM_reduced} is not closed and, therefore, cannot be used to propagate the reduced density matrix in the present form.
This fact can also be regarded as a consequence of the semi-reduced nature of $\mathbb{V}$ retaining the $m$-dependence through the angular momentum part of the \ac{SOC} operator, Eq.~\eqref{eq:V_SOC}.
Moreover, the submatrices $\Vmat^m$ of the semi-reduced \ac{SOC} Hamiltonian couple $\smmat{k\,q}$ to $\smmat{k\,(q+m)}$, meaning that, to account for \ac{SOC}, for a given $k$, the $q=\pm1$ elements require the $q=\pm2$ projections to be propagated, which, in turn, depend on $q=\pm3$ and so forth. 
As this iteratively leads to full propagation, 
one does not benefit from the reduction.

Nevertheless, the structure of Eq.~\eqref{eq:EOM_reduced} can inspire physically motivated rank and component truncation in the full \ac{EOM}, Eq.~\eqref{eq:LvN_spherical}. 
First, only the lowest $\pm q$ projections for each rank $k$ (or $K$) can be considered the most important, e.g., $q=0,\pm1$.
Second, the ranks themselves can be truncated at some value below the maximum $2\cdot\max\{S,S',S''\}$, thus, neglecting all of the computationally demanding higher-rank contributions.
In this case, the equations for the truncated expansion are closed, providing substantial savings in the computational effort.

\subsection{Implementation details}
\label{subsec:implementation}

\begin{figure*}[tb!]
    \centering
    \includegraphics[width=0.75\textwidth]{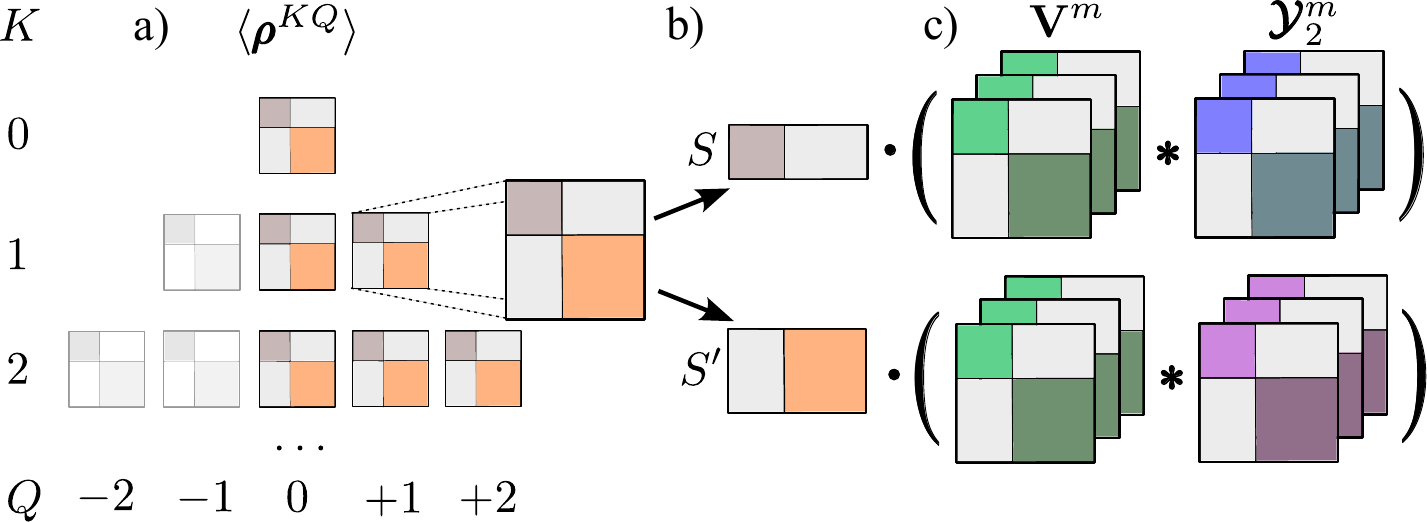}
    \caption{Computational scheme for evaluating \ac{SOC}-dependent terms in the right-hand side of \ac{EOM} employing matrix-matrix multiplication, Eq.~\eqref{eq:LvN_spherical}. a) only part of elements of $\langle \pmb{\rho}^{K\,Q}\rangle$ is computed due to hermiticity;
    b) for every $K\,Q$, the state multipole (density) matrix is split into slabs with the same spin in rows;
    c) every slab is matrix-matrix (denoted as $\cdot$) multiplied with the Hadamar products (denoted as $\ast$) of $\mathbf{V}^m$ and respective $\pmb{\mathcal{Y}}_i^m$. 
    }
    \label{fig:comp_scheme}
\end{figure*}

Obtaining wave functions of states $\ket{aSM}$ and matrix elements of the $\hat H_\text{CI}$, $\hat {\pmb{\mu}}$, and $\Vop$ operators was performed with the \texttt{OpenMolcas} code.~\cite{LiManni_JCTC_2023}
The propagation itself was performed with the \texttt{RhoDyn} module of the same package,~\cite{Kochetov_JCTC_2022} which has been extended to allow for the decomposition and time propagation of the density matrix in the spherical tensor basis.

Particular emphasis should be put on accurately evaluating the numerical value on the right-hand side of Eq.~\eqref{eq:LvN_spherical}, namely the sums over the $\left\{\mathbb{V}, \smmat{KQ}\right\}$ \ac{SOC} terms, see Fig.~\ref{fig:comp_scheme}.
Element-wise propagation of each matrix element of the state multipoles is not efficient. 
It is replaced by the propagation of $\smmat{k\,q}$ matrices as a whole and involves matrix-matrix multiplications, which are efficiently realized in mathematical libraries.
However, incorporating such a multiplication is more complex due to 3j-symbols in $\mathcal{Y}_1$ and $\mathcal{Y}_2$ coupling different ranks.

To circumvent this complication, we first note that geometrical factors $\mathcal{Y}_i$ depend on ranks and projections $k, q, K, Q$, as well as on spins $S,S',S''$ and $m$.
Therefore, the computation must be performed separately for every  $k, q, K$, and $Q$ combination.
When fixing $S'$ (or $S$ for the second term) and, thus, splitting the $\smmat{KQ}$ into slabs, Fig.~\ref{fig:comp_scheme}b), one can compute matrices $\pmb{\mathcal{Y}}_1^m$ ($\pmb{\mathcal{Y}}_2^m$) depending only on the number $m$, see Eqs.~\eqref{eq:ypsilon1}~and~\eqref{eq:ypsilon2}.
Now, Hadamard products $\pmb{\mathcal{Y}}_i^m \ast \pmb{V}^m$ can be computed as schematically shown in Fig.~\ref{fig:comp_scheme}c);
the result is multiplied by a slab of $\smmat{K\,Q}$, and the resulting slabs are assembled into the full matrices contributing to the given $\smmat{k\,q}$ derivative.

In addition, one can utilize the hermiticity of the density matrices $\smmat{k\,-q} = (-1)^{S-S'-q}\smmat{k\,q}^\ast$.
Therefore, either $\smmat{k\,q}$ or $\smmat{k\,-q}$ is calculated using Eq.~\eqref{eq:LvN_spherical}, and the other can be easily obtained.
In the discussion below, we will denote both $\smmat{k\,\pm q}$ as $\smmat{k\,q}$ unless specified otherwise. 
The algorithm, involving matrix multiplication and accounting for hermiticity, substantially speeds up the calculations.

\subsection{Computational details}
\label{subsec:compdetails}

The \Ti\ molecular geometry, a tetragonal structure of $T_d$ point symmetry with Ti--Cl distances of 2.170~\AA, was taken from Ref.~\cite{Morino_JCP_1966} and was obtained by gas electron diffraction.
The \FeWat\ and \CrWat\ structures were obtained at the DFT level with the B3LYP functional and aug-cc-pVTZ basis set in the \texttt{Gaussian} program package;~\cite{Frisch2016} see Ref.~\cite{Kochetov_JCP_2020} for more details.
Hexaaqua complexes possess approximate $O_h$ symmetry, lowered by the presence of H atoms and the Jahn-Teller effect in \FeWat.
Metal-oxygen distances are 2.04, 2.27~\AA\ for \FeWat, and 2.00~\AA\ for \CrWat.

Calculations of \ac{SF} states and interstate couplings are performed at the \ac{RASSCF} level of theory.
Scalar relativistic effects are accounted for via a Douglas-Kroll-Hess transformation~\cite{Douglas_AP_1974} up to the second order within the perturbation theory framework.
The ANO-RCC basis set of TZ quality is used for all atoms.
The active space of 8 orbitals (three $2p$ and five $3d$ orbitals of transition metals) gave a good approximation for the core-excited states of ionic complexes~\cite{Bokarev_WCMS_2020} and is used for all species.
Full-\ac{CI} has been done for the $3d$ subspace (RAS2), while only one hole has been allowed for the $2p$ subspace (RAS1).
The RAS3 subspace has been left empty, apart from \Ti; see Sec.~\ref{subsec:ti}.
Quantities needed for propagation are obtained from the \texttt{h5} output file from the \texttt{OpenMolcas} \ac{RASSI} module.
\ac{SOC} matrix elements are computed by making use of \acl{AMFI}.~\cite{Schimmelpfennig1996}
Propagation of the state multipoles according to Eq.~\eqref{eq:LvN_spherical} was performed by the Runge-Kutta method of the fourth order.

For simplicity, the incoming electric field has been chosen to be a single linearly polarized pulse with a temporal Gaussian envelope
\begin{equation}\label{eq:pulse}
    \pmb{\mathcal{E}}(t) = A\,\vec{e}\,\exp{(-{(t-t_0)^2}/({2\sigma^2}))}\sin (\Omega t)\ ,
\end{equation}
where $A$, $\vec e$, $t_0$, and $\Omega$ are the amplitude, polarization, center of the envelope, and carrier frequency.
In all cases, the pulse envelope is centered at $t_0=0.5\,$fs and has $\sigma = 0.125\,$fs.
The pulse width $\sigma$ has been chosen to cover a wide range of valence-core excitations;
thus, it corresponds to the ultrashort pulse in the time domain.
The carrier frequency $\Omega$ has been chosen for each case to be centered in the middle of the \Ledge\ absorption edge in the frequency domain and to cover all bright transitions.
Dynamics have been simulated in a time interval of 3\,fs which is enough to see the main features and is slightly smaller than the typical $2p$ core-hole lifetime.

\section{Results and Discussion}
\label{sec:Results}

We consider three cases from simplest to more complicated.
In the \Ti\ molecule, one has a small number of participating singlet ($S=0$) and triplet ($S=1$) states.
Although this case is quite instructive in understanding the internal symmetries of the state multipoles $\smmat{k\,q}$ and their physical meaning, it does not exhaust other, more interesting situations.
\FeWat\ is a high spin complex with the quintet ($S=2$) ground state, where we have considered the coupling of the quintet manifold to the triplet ($S=1$) one.
\CrWat\ is a further example where the coupling of three different manifolds -- initially populated quartet ($S=3/2$) to doublet ($S=1/2$) and sextet ($S=5/2$) -- is regarded.
These two cases are more illustrative when it deals with truncating ranks and projections of spherical tensors due to a much larger number of participating states than in the case of \Ti.

Further, we will distinguish two initial conditions representing:
i) equal population of $\pm M$ components for a given $M$, e.g., according to Boltzmann equilibrium distribution at finite temperature $T${=273\,K};
ii) asymmetric populations, e.g., when a single component (either $+M$ or $-M$) is populated, whereas the other is not.
The latter case corresponds to breaking the time-reversal symmetry, e.g., due to external magnetic fields or 'spin filtering' with some kind of Stern-Gerlach experiment.
It should be stressed that we populate an initial density matrix in the state basis and only then perform the expansion of it into a series of state multipoles $\smmat{k\,q}$ according to Eq.~\eqref{eq:tens_exp}.

Below, we restrict ourselves to analyzing solely the diagonal elements of the density matrix in the spherical tensor basis.
These elements are directly connected to state populations {and coherences between them}, see Appendix~\ref{app:meaning}.
Absolute values of diagonal elements are summed over diagonal subblocks ($\sum_{a}|\langle\rho^{k\,q}_{aS,aS}\rangle |$) with a certain multiplicity.
They are denoted as $X^{kq}$, where $X$ stands for Q (quintets), T (triplets), S (singlets) for integer spins and S (sextets), Q (quartets), and D (doublets) for half-integer spins. 

\subsection{{Highly coherent case: \Ti\ system}}
\label{subsec:ti}

\begin{figure}[tb!]
	\centering
	\includegraphics[width=0.48\textwidth]{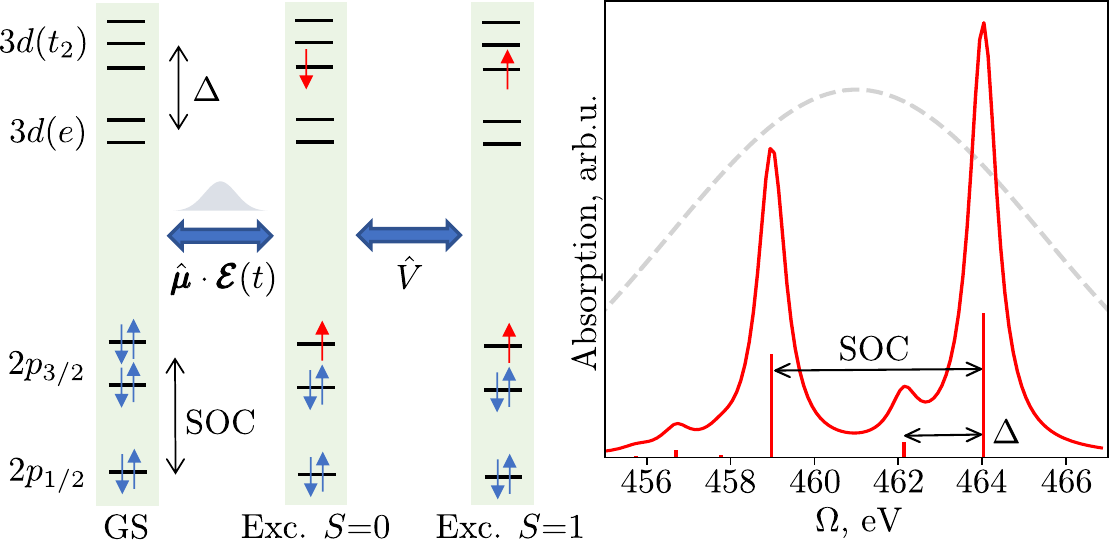}
	\caption{
	(Left) Schematic depiction of the participating states for the \Ti\ system and the couplings between these states due to external field and \ac{SOC}; (Right) X-ray absorption spectrum, red line, and the envelope of the Fourier transformed pulse, dashed gray line. In the \ac{MO} active space, the orbitals that are depicted close together are degenerate; $\Delta$ corresponds to the orbital ligand-field splitting (left).}
	\label{fig:ti_scheme}
\end{figure}

For the \ac{RASSCF} calculations of \Ti, three $2p$ orbitals of titanium make up the RAS1 subspace, and five $3d$ orbitals make up the RAS3 subspace, while RAS2 is left empty.
For RAS1, one hole is allowed, and for RAS3, one excited electron is allowed that corresponds to a single core excitation to the empty $3d$ orbitals.
Thus, the setup is equivalent to the CI-Singles level of theory.
With such a setup, 16 singlet and 15 triplet states are obtained, leading to 31 spin-free and 61 basis spin states.
The pulse that triggers the spin dynamics has the form of Eq.~\eqref{eq:pulse}, with parameters $A=1.5$\, a.u. and $\Omega = 461\,$eV.
It is chosen to be centered in the middle of the \Ledge\ absorption edge in the frequency domain and to cover all states (see Fig.~\ref{fig:ti_scheme}).
The full density matrix for the \Ti\ molecule was propagated both according to Eq.~\eqref{eq:LvN_spherical}, where each matrix $\smmat{k\,q}$ has the dimensions of 31$\times$31 \ac{SF} states, as well as using the conventional approach of LvN Eq.~\eqref{eq:LvN} in the basis of 61 spin states.

\begin{figure}[tb]
	\centering
	\includegraphics[width=0.48\textwidth]{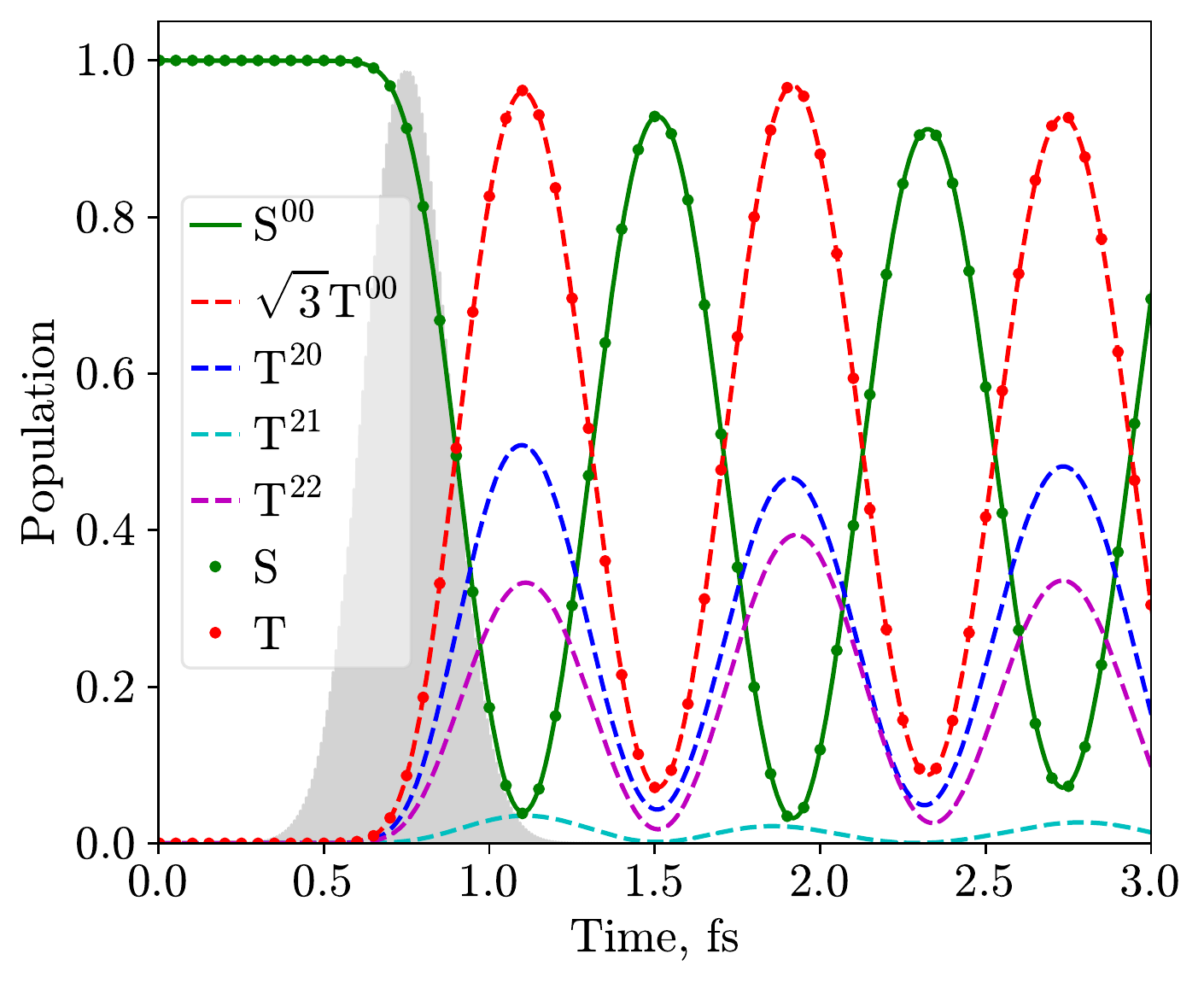}
	\caption{\Ti: The evolution of the summed diagonal elements ($\sum_{a}|\langle\rho^{k\,q}_{aS,aS}\rangle |$ for a given spin $S$) of state multipoles according to Eq.~\eqref{eq:LvN_spherical} (lines) and populations in the spin-state basis from the conventional dynamics (points). 
		Singlet (S) states are denoted with solid line, while triplets (T) -- with dashed lines.
		T$^{00}$ is scaled with $\sqrt{3}$ to obtain populations, see Eq.~\eqref{eq:diagK0}.
	}
	\label{fig:ti_dyn}
\end{figure}

The X-ray absorption spectrum of \Ti\ is displayed in Fig.~\ref{fig:ti_scheme}, and the principal scheme demonstrates the electronic structure of involved states.
The spectrum represents the two groups of bands corresponding to $2p_{3/2}$ and $2p_{1/2}$ hole states split by strong \ac{SOC}.
There is also a smaller splitting $\Delta$ due to the tetrahedral field of the Cl atoms.
The pulse is broad enough to populate all the bright states.
In the \ac{SF} picture, these are two bright singlet core-excited states having a multiconfigurational character and consisting of singly-excited configurations $(2p)^{-1}(t_2)^1$ and $(2p)^{-1}(e)^1$.
By $t_2$ and $e$, we denote manifolds of virtual valence orbitals of mainly Ti $3d$ character.
The intensity of the peak with a primary $t_2$ contribution is much higher than the other one with a primary $e$ contribution.~\cite{Decleva_CP_1994,Stener_CPL_2003}
The pulse characteristics are chosen to deplete the ground state  within the pulse duration almost completely.
\ac{SOC} couples the initially excited singlet states to a handful of triplet states, whereas almost two-thirds of triplet states do not participate in the dynamics, see the analysis in Ref.~\cite{Kochetov_JCP_2020} for a similar system.
This coupling drives the transition of the population from singlet to triplet manifolds leading to quantum beating similar to Rabi oscillations, Fig.~\ref{fig:ti_dyn}.
Due to the relatively small number of participating states, these oscillations correspond to an almost complete population transfer between manifolds.
For the same reason, these oscillations do not decay, and the dynamics remain in a highly coherent regime, see discussion in Sec.~\ref{subsec:fe_cr}.

Fig.~\ref{fig:ti_dyn} compares the collective population of singlet and triplet states computed in the basis of 61 spin states (red and green dots) with the evolution of different state multipoles.
S$^{00}$ and T$^{00}$ correspond to the sums of diagonal elements of the singlet-singlet and triplet-triplet blocks of $\smmat{0\,0}$;
when scaled with the square root of multiplicity, T$^{00}$ is equivalent to the overall population of the triplet manifold in a state basis.
The period of the oscillation between the singlet and triplet states is roughly 0.82\,fs, which corresponds to an energy of 5.04\,eV that is in accord with the \ac{SOC} splitting between $2p_{1/2}$ and $2p_{3/2}$ hole states of 5.05~eV, Fig.~\ref{fig:ti_scheme}. 
The results in the basis of spherical tensors and states coincide, which serves as an essential internal consistency check.

For singlet states, the S$^{kq}$ are strictly zero for $k>0$.
For the triplet-triplet block, higher ranks are also possible.
However, the evolution of $\smmat{1\,q}$ is not shown in Fig.~\ref{fig:ti_dyn}, as these state multipoles always stay zero, even for triplet states. The $q=0$ component for $k=1$ characterizes the asymmetry in the population of $M$ and $-M$ states, i.e., the system's polarization which stays zero because of the time-reversal symmetry (no magnetic interactions) and initial conditions.
Respectively, the diagonal elements of the $q=\pm1$ projections, characterizing the coherences between $M$ and $M\pm1$ projections of the same basis state, stay strictly zero, see Eq.~\ref{eq:diaK1Q1}. 
The diagonal $\smmat{2\,2}$ describes coherences between states with $\Delta M = \pm 2$, i.e., $M=1$ and $M=-1$, where the direct transition is not possible.
Nevertheless, they can be simultaneously populated or depopulated, mediated by the intermediate singlet state with $M=0$.
The $\smmat{2\,1}$ corresponds to a similar process but reflects the coherence between $M=\pm1$ and $M=0$, mediated by some other state.
Note that in contrast to $\smmat{1\,1}$, $\smmat{2\,1}$ does not break the time-reversal symmetry.
The off-diagonal elements correspond to coherences and are not easy to analyze.
Since we are interested in the population dynamics, they will not be further discussed.

In general, the applicability of the method has been tested for \Ti.
The flexibility of the density matrix approach allows one to describe this highly coherent case exhibiting distinct quantum beatings.
However, in this case, the reduced dynamical description with limited highest ranks and projections is impossible as discussed further in Sec.~\ref{subsec:red_prop}.
Below we consider less coherent cases where the larger number of involved states, serving as the discrete electronic ``reservoir'', leads to the (reversible) phase relaxation.

\subsection{Less coherent cases: \FeWat\ and \CrWat}
\label{subsec:fe_cr}

\begin{figure}[tb]
	\centering
	\includegraphics[width=0.48\textwidth]{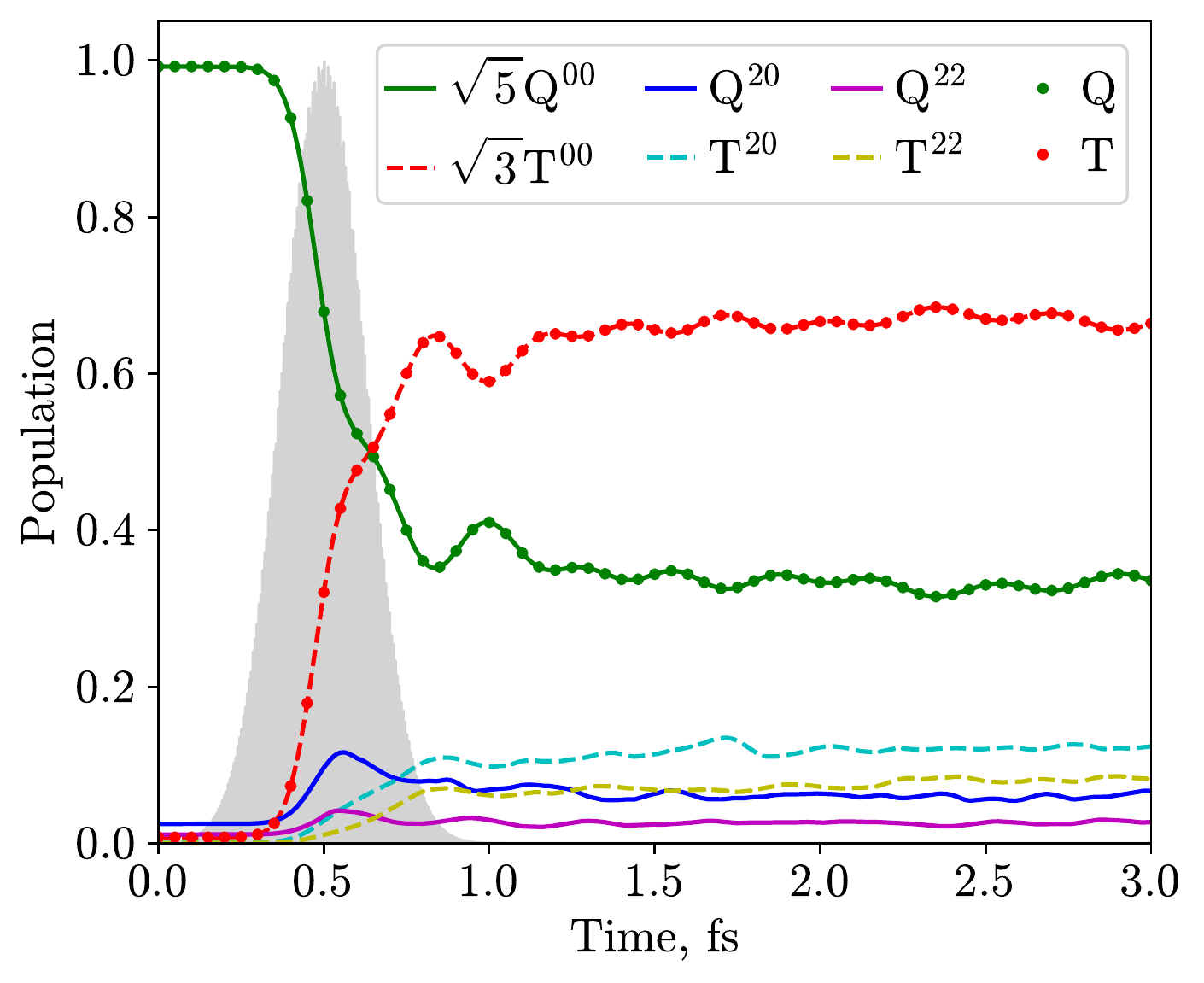}
	\caption{\FeWat: The evolution of summed diagonal elements of state multipoles (solid and dashed lines) and diagonal elements of the density matrix in the spin-state basis (dots).
	The initial populations are according to equilibrium distribution at $T=273$\,K. 
	State multipoles with $k>2$ are not shown because of their smallness.
	To estimate the magnitude of their contributions, see Fig.~\ref{fig:fe_kq_hist}(a).
	}
	\label{fig:fe_dyn_so}
\end{figure}

Next, we consider two less coherent cases -- spin dynamics in hexaaqua Fe$^{2+}$ and Cr$^{3+}$ complexes.
\FeWat\ has a $d^6$ electronic configuration and a high-spin quintet ($S=2$) ground state.
The full \ac{SF} basis in the $2p3d$ active space, see Sec.~\ref{subsec:compdetails}, constitutes 35 quintet and 195 triplet states.
Upon account for the spin part, a natural state basis consists of 760 states.
The maximum rank of the state multipoles $k_\text{max}=4$ is required to represent the quintet--quintet block of the density matrix exactly and to reproduce the dynamics in the state basis.
The evolution of populations in the spin-state basis and diagonal state multipoles for quintets (Q) and triplets (T) is depicted in Fig.~\ref{fig:fe_dyn_so} with points and lines, respectively.
Pulse characteristics used in this case are $A=6$\, a.u. and $\Omega = 712\,$eV; see Eq.~\eqref{eq:pulse}.

\begin{figure}[tb]
	\centering
	\includegraphics[width=0.48\textwidth]{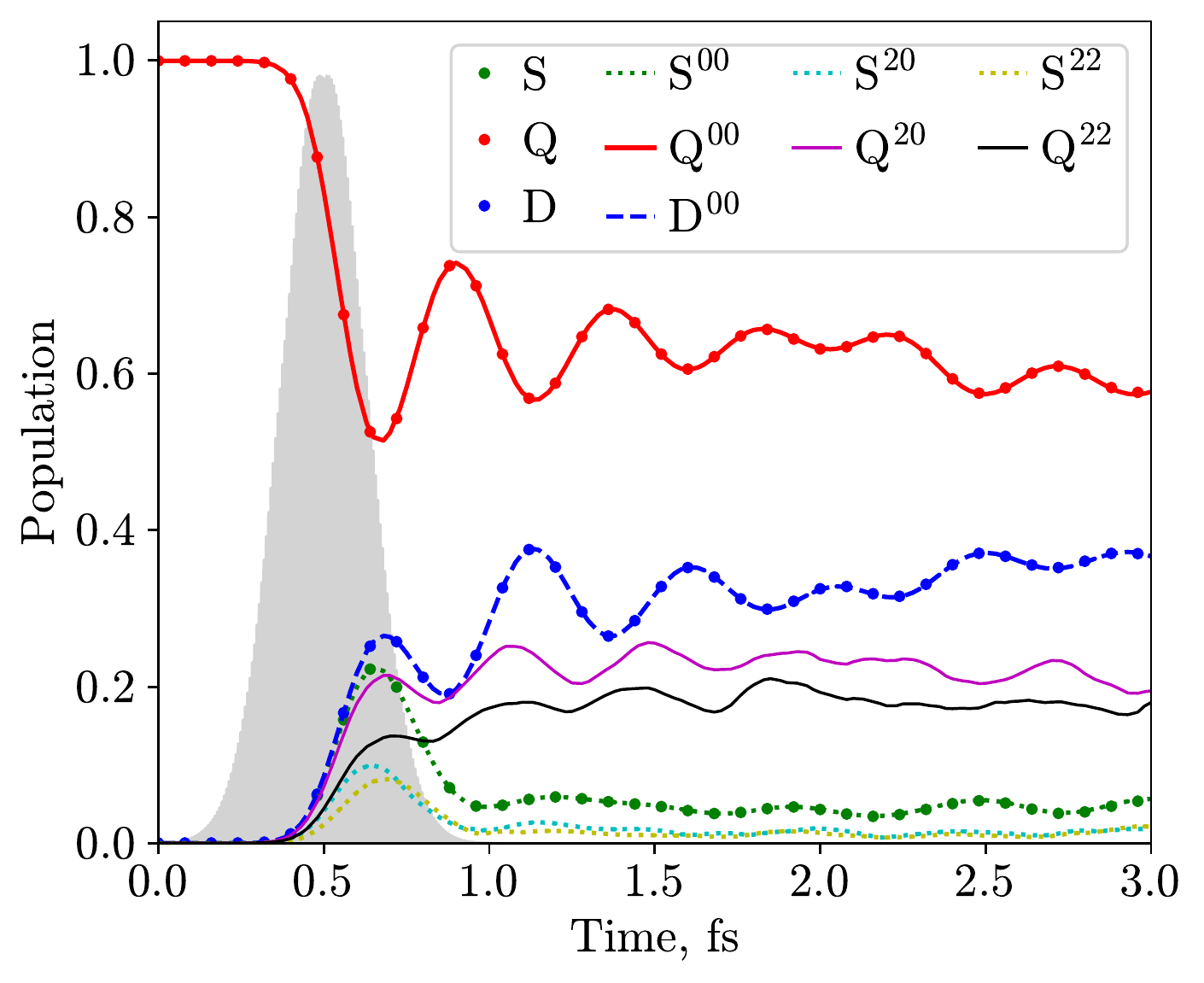}
	\caption{
	\CrWat: Evolution of diagonal elements of the state multipoles for the initial equilibrium distribution at $T=273\,$K of the initial density matrix.
	Quartet (Q) states are denoted with solid lines, while doublets (D) -- with dashed lines, and sextets (S) with dotted lines.
	The results of the propagation in the state basis are displayed with points.
	State multipoles with $k>2$ are not shown.
	}
	\label{fig:cr_dyn}
\end{figure}

The other system, \CrWat, has a richer electronic structure regarding the number of participating states and different multiplicities.
In the $2p3d$ active space, one has in total 15  sextet, 160  quartet (including the ground state), and 325  doublet \ac{SF} states giving rise to 90, 640, and 650 spin states, respectively.
Thus, the dimensions of the SF and spin bases are 500 and 1380.
Pulse characteristics used for the \CrWat\ complex in Eq.~\eqref{eq:pulse} are $A=2.5$\,a.u. and $\Omega = 588\,$eV.
The population of state multipoles can be inferred from Fig.~\ref{fig:cr_dyn}.

Again, as for \Ti, approximately only half of the total number of states are notably participating in dynamics for both \FeWat\ and \CrWat\ complexes, but in the latter cases, their number is at least an order of magnitude larger than for \Ti.  
This fact determines the different character of dynamics: on the one hand, one also sees a fast rise of the populations of the flipped-spin states, e.g., triplets in the case of \FeWat, see Fig.~\ref{fig:fe_dyn_so}.
On the other hand, oscillations are less prominent and generally tend to decay about 1\,fs after the pulse.
One can consider it an ``equilibration'' in a discrete quasi-reservoir of electronic states.
However, this dephasing is reversible, and a rephasing should happen later as the system is closed.

It is not easy to quantify the degree of coherence for the considered cases.
As some conditional quantification, we have computed the Shannon entropy, mean (averaged over all $N(N-1)/2$ elements) time-averaged absolute value of the off-diagonal elements of the density matrix (analog of the $||\pmb{\rho}||_1$ norm), and maximum absolute off-diagonal element (analog of the $||\pmb{\rho}||_\infty$ norm).
For \Ti, the entropy stays around 0 during the propagation, the mean off-diagonal element is around 3$\cdot 10^{-3}$, and the maximum off-diagonal element is 0.3.
In contrast, for \FeWat\ assuming thermal population, the entropy is around 1.75, the mean coherence is an order of magnitude smaller around 2.5$\cdot 10^{-4}$, as the maximum coherence is 0.05.
These diagnostics show that the case of \FeWat\ is much less coherent than \Ti.
The reasons are the initial incoherent population of states according to Boltzmann distribution and the larger number of participating states acting as quasi-reservoir.

\begin{figure}[tb]
	\centering
	\includegraphics[width=0.48\textwidth]{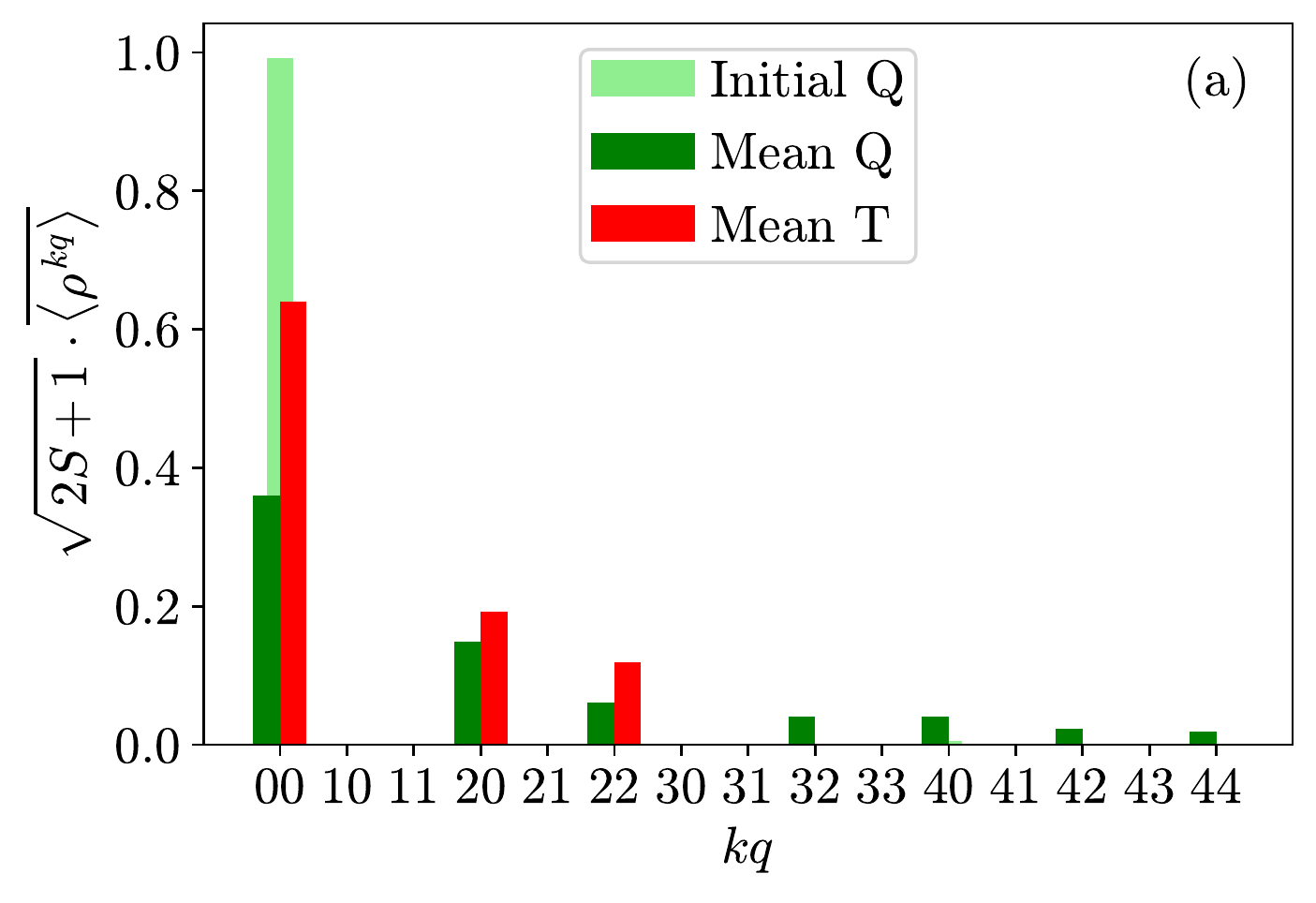}\\
	\includegraphics[width=0.48\textwidth]{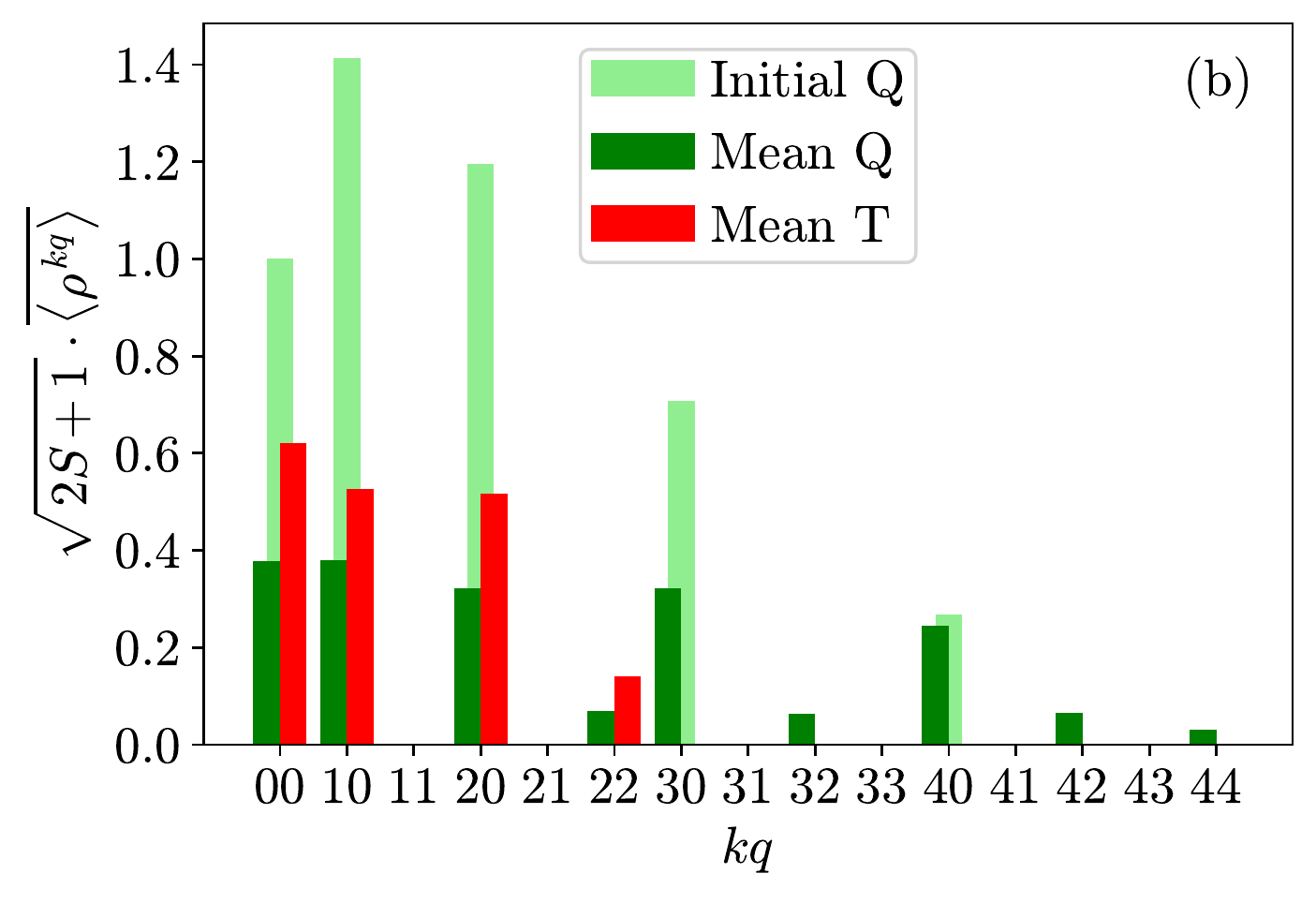}
	\caption{\FeWat: Mean contributions (i.e., absolute values of diagonal elements summed according to spin multiplicity and averaged over time) of state multipoles to the density matrix.
	Averaging over time is performed from the center of the pulse till the endpoint at 3\,fs.
	The initial density matrix 
	         (a) is populated according to equilibrium distribution at  $T=273$\,K;
	         (b) includes a total population in a single $M$ component of the ground state.
	         The decomposition of $\pmb\rho$ values at $t=0$~fs is denoted with light-green bars.
	         All values are scaled with the factor $\sqrt{2S+1}$ such that 00 contributions correspond to populations.}
	\label{fig:fe_kq_hist}
\end{figure}

\begin{figure}[tb]
	\includegraphics[width=0.48\textwidth]{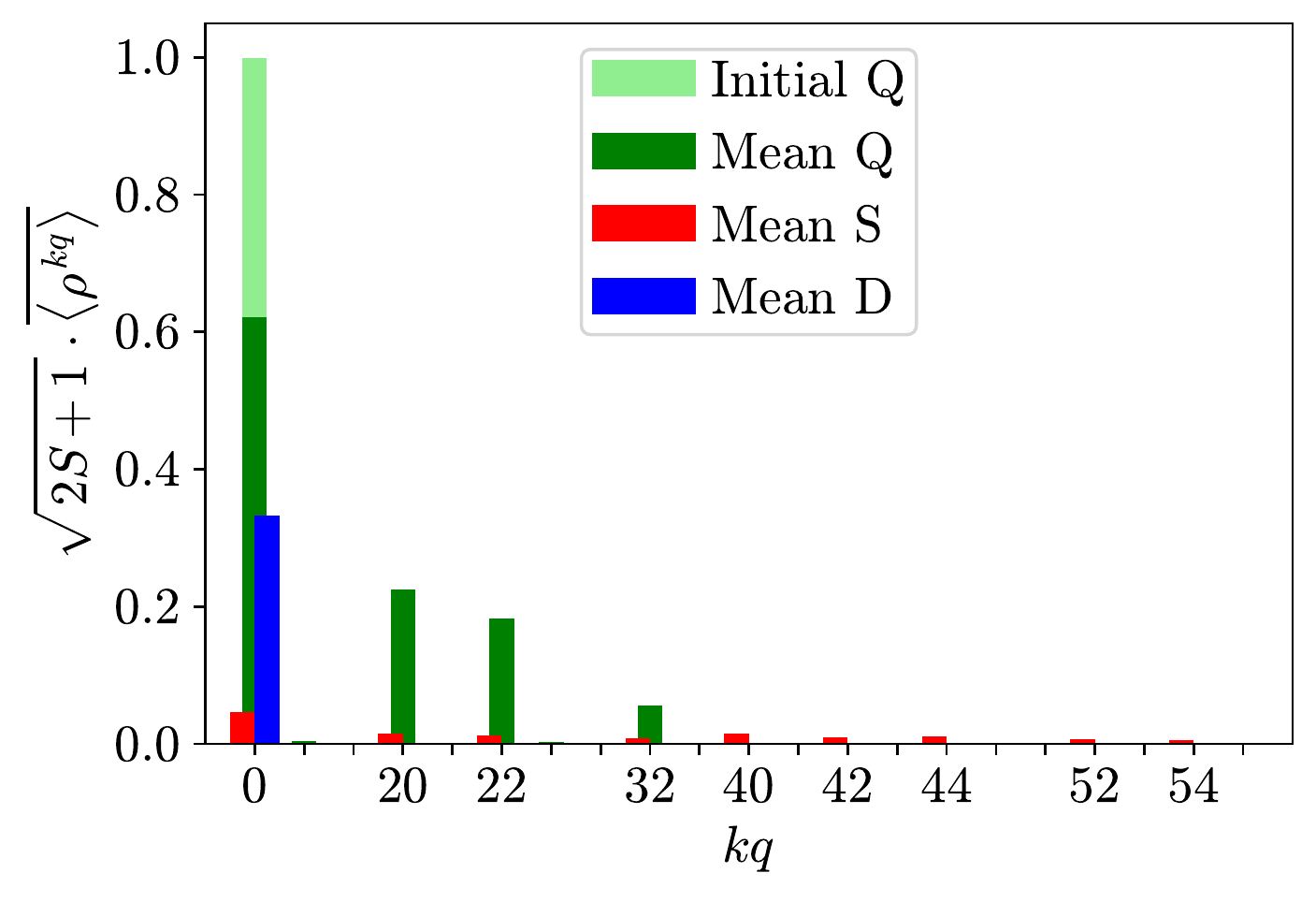}
	\caption{
	\CrWat: Mean contributions of the state multipoles to the density matrix for the equilibrium initial population at $T=273$\,K.
	See also the caption to Fig.~\ref{fig:fe_kq_hist}.
	}
	\label{fig:cr_hist}
\end{figure}

Illustrative is the participation of different state multipoles in the dynamics.
It is analyzed here for the diagonal contributions summed over different diagonal subblocks of the density matrix, see, e.g., Figs.~\ref{fig:fe_dyn_so} and~\ref{fig:cr_dyn} for the time evolution and Figs.~\ref{fig:fe_kq_hist}(a) and~\ref{fig:cr_hist} for the decomposition of the initial density matrix and the contributions averaged over the full simulation time for \FeWat\ and \CrWat.
The first thing to notice is that for the equilibrium population of the initial states, a relatively small number of $k\,q$ components contribute.
For instance, in Fig.~\ref{fig:fe_dyn_so}, only contributions substantially different from zero are plotted, namely, $k=0$ and $k=2$ with even projections $q=0,2$ for both quintet and triplet states.
Others are either strictly zero or rather small.
Fig.~\ref{fig:fe_kq_hist}(a) further illustrates this observation.
One sees that the decomposition of the initial density matrix according to Eq.~\eqref{eq:tens_exp} (light green bar) involves only $\smmat{0\,0}$ multipoles.
In the course of dynamics, the system is in a superposition of triplet (red) and quintet (green) excited states with a population ratio of about 3:2.

A similar situation is observed for \CrWat, Fig.~\ref{fig:cr_hist}.
The importance of higher ranks quickly decreases, with doublets being restricted only to the first rank, quartets to the third rank, and sextets going to the highest fifth rank.

To conclude on these cases, the uniform distribution of the initial population is followed by consecutive ``uniform'' dynamics, mainly involving the low-rank state multipoles.
The analysis of diagonal contributions for less coherent cases shows that one can cut not only the ranks but their components as well.
This builds a basis for reduced propagation, see Sec.~\ref{subsec:red_prop}.

To explore the situation when the initial population of $\pm M$-microstates is uneven, e.g., due to interaction with the magnetic field, we populate only a single ground-state $M$-component.
It has to be noted that the overall dynamics in the state basis for equilibrium and non-equilibrium initial conditions are almost identical, as was shown in our previous work.~\cite{Kochetov_JCP_2020}
However, one can discern variations in the evolution of spin-state components with distinct spin projections by analysis of state multipoles.
From this viewpoint, the non-equilibrium initial condition leads to a different scenario.
The situation is notably more coherent as the entropy stays around 0 during the propagation, the mean coherence is around 3.5$\cdot 10^{-4}$, and the maximum coherence is 0.25, comparable to that of \Ti. It can be rationalized by a smaller number of initially excited states because the coupling to the electromagnetic field conserves $M$, Eq.~\ref{eq:muE}.
The mean contributions of the state multipoles are shown in Fig.~\ref{fig:fe_kq_hist}(b).
The initial population distribution, which is asymmetric for different $M$-projections, involves all ranks of the quintet subblocks, $\smmat{k\,0}$ (light green bars).
This leads to more diverse {population redistribution} between different ranks during the dynamics.
For instance, contributions from the $\smmat{1\,0}$ and $\smmat{3\,0}$ multipoles become non-zero, evidencing a significant degree of asymmetry between $M$ and $-M$ states.
In other words, such non-equilibrium initial conditions offer fewer possibilities for rank and projection truncation, as discussed in Sec.~\ref{subsec:red_prop}, since higher ranks and projection play a more critical role.

\begin{figure}[h!]
\centering
\includegraphics[width=0.48\textwidth]{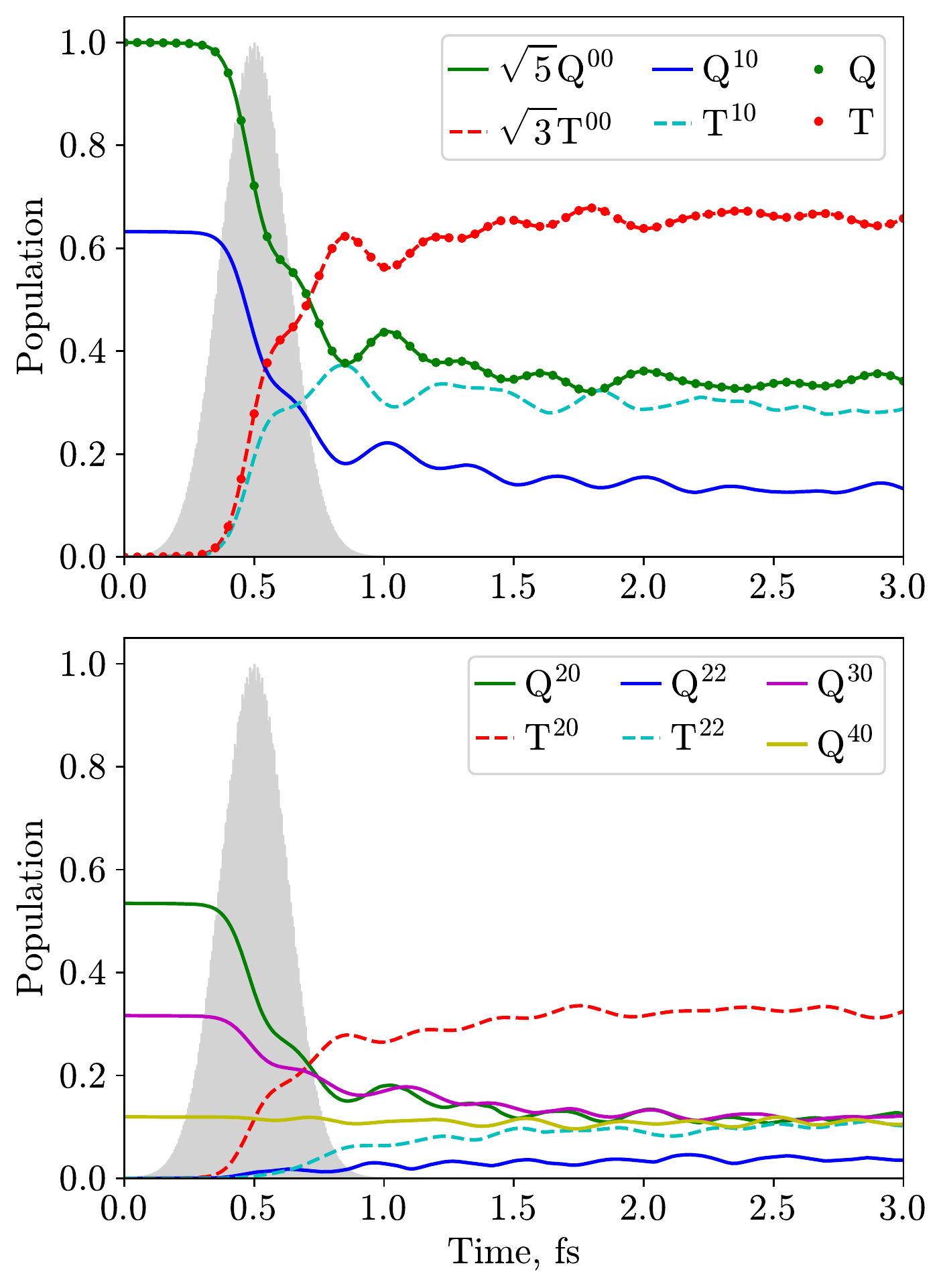}
\caption{\FeWat: Evolution of diagonal elements of state multipoles and the populations obtained by performing dynamics in the state basis (dots);
    see also caption to Fig.~\ref{fig:fe_dyn_so}.
	Only a single $M$-component of the ground state is initially populated.
	The non-zero components Q$^{32}$, Q$^{42}$, and Q$^{44}$ are not shown; their magnitude can be estimated from Fig.~\ref{fig:fe_kq_hist}(b).
}
\label{fig:fe_dyn_sf}
\end{figure}

\subsection{Reduced Propagation}
\label{subsec:red_prop}

\begin{figure}
	\centering
	\includegraphics[width=0.48\textwidth]{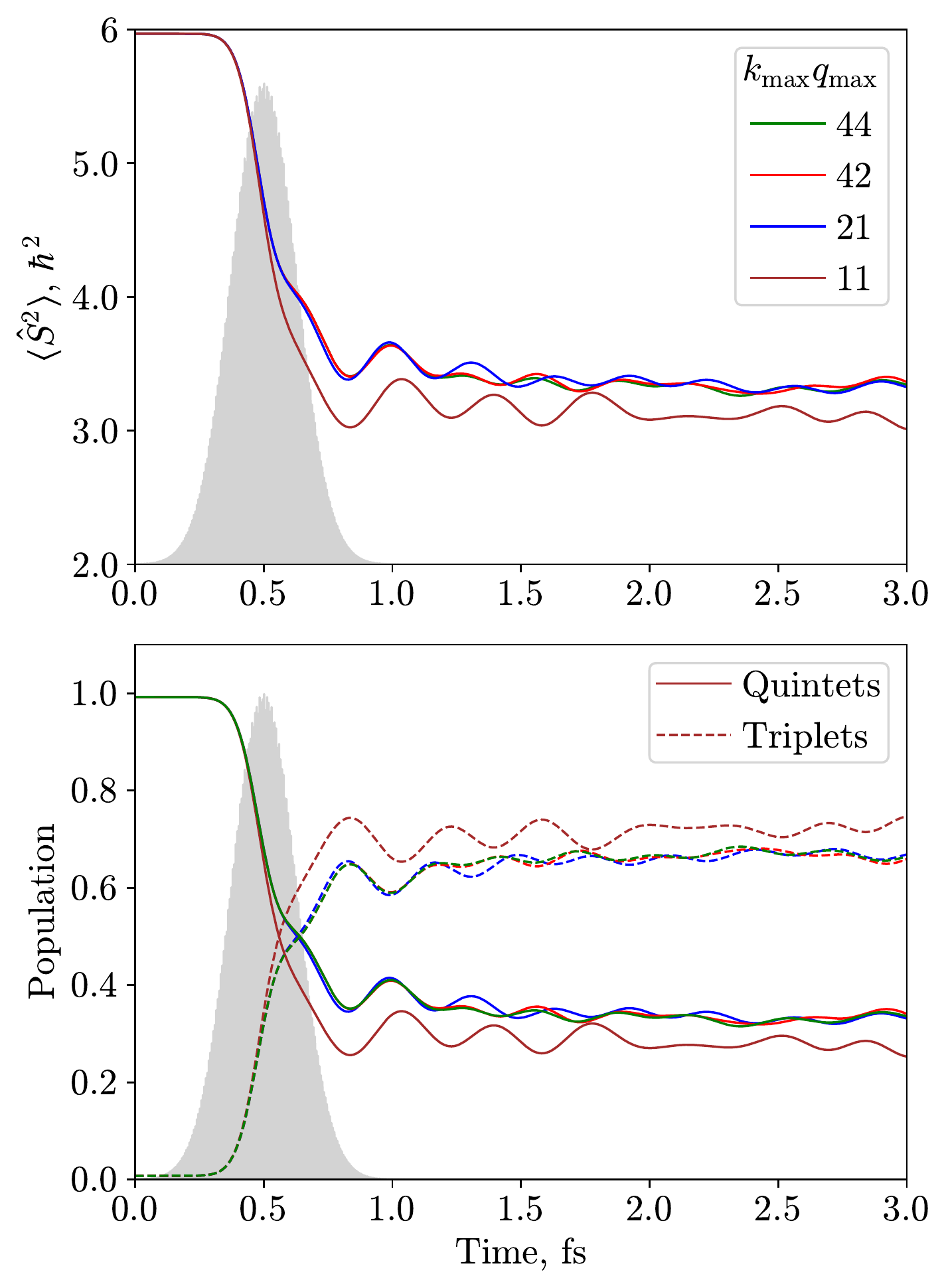}
	\caption{
	\FeWat: Evolution of \Ssq\ (upper panel) and populations (lower panel) for the full $k_\text{max}\, q_\text{max}=4\,4$, and truncated propagation.
	Quintet states' population is denoted with solid lines, while triplet -- with dashed lines.
	The initial population corresponds to $T=273\,K$.
	}
	\label{fig:fe_dyn_truncated}
\end{figure}

\begin{figure}
	\centering
	\includegraphics[width=0.48\textwidth]{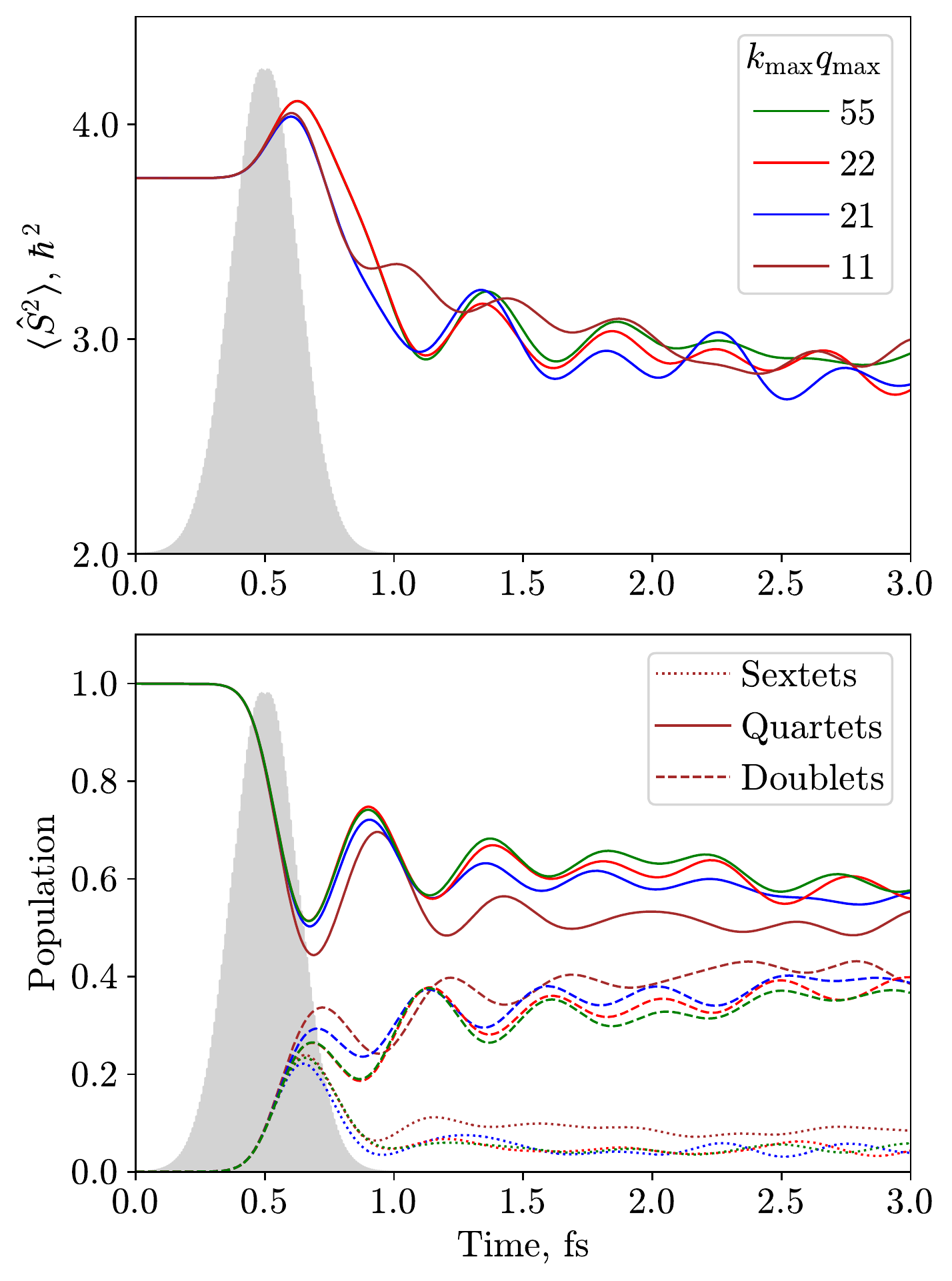}
	\caption{\CrWat: Evolution of \Ssq\ (upper panel) and populations (lower panel) for the full $k_\text{max}\, q_\text{max}=5\,5$, and truncated propagation.
	The initial population corresponds to $T=273$\,K.
	}
	\label{fig:cr_dyn_truncated}
\end{figure}

This section discusses the results of truncating the state multipole expansion at certain $k$ and $q$ below maximum.
We will call the respective dynamics ``reduced''.
Nevertheless, it is obtained using the full equation, Eq.~\eqref{eq:LvN_spherical}, but not the reduced one, Eq.~\eqref{eq:EOM_reduced}, as the latter is not closed.
The reduced \ac{EOM} is used only to guide the truncation.
{The truncation is performed simply by setting all multipoles for $k$ and $q$ larger than the threshold value to zero and constraining the sum over $K$ and $Q$ in Eq.~\eqref{eq:LvN_spherical}.}
The results of the reduced propagation are displayed in Figs.~\ref{fig:fe_dyn_truncated} and~\ref{fig:cr_dyn_truncated} of \FeWat\ and \CrWat, respectively.
From these figures, one can also infer the time-dependent expectation value of the spin-squared operator $\langle \hat S^2\rangle$.

First, we note that the projections for $k>0$ should include at least $\pm1$, which follows from the reduced \ac{EOM}, Eq.~\ref{eq:EOM_reduced}.
Indeed, although the dynamics including only $\smmat{k\,0}$ (not shown) also predicts spin flip, they substantially differ from the exact.
There must be more than just $k=1$ terms, as $k=2$ also notably contributes.
The truncation at $k=2$ and $q=1$ is sufficient to  reproduce the dynamics semiquantitatively.
Although from Figs.~\ref{fig:fe_kq_hist}(a) and~\ref{fig:cr_hist}, it seems logical to cut at $k\,q=2\,0$ or at $2\,2$ both for \FeWat\ and \CrWat, one should keep in mind that shown are only the diagonal elements that do not account for essential coherences.
The latter are sufficiently represented by the $\smmat{1\,1}$ and $\smmat{2\,1}$ multipoles.
The \CrWat\ case is especially illustrative regarding truncation since the higher ranks up to $k=5$ seem formally important.
They correspond to a relatively small number of sextet states and can be neglected, leading to substantial savings in the effort.

Importantly, this truncation is most prominent for less coherent cases. E.g., in \FeWat\ with non-equilibrium initial population, the ranks cannot be truncated. 
Only minor savings in computational time can be achieved by excluding some projections for higher ranks.
Further, the reduction is impossible for the highly coherent example of the \Ti\ molecule, as the truncation at $k=2, q=1$ produces wrong results (not shown).
The $\smmat{2\,2}$ component describes critical coherent pathways governing the population redistribution that cannot be neglected.

The observation of the decisive role of coherence and a typical truncation at $k\,q=2\,1$ can be rationalized as follows:
The entirely incoherent dynamics can be represented solely by the $\smmat{0\,0}$ as it governs the diagonal of the density matrix, and off-diagonal elements are zeros.
With increasing coherence, higher ranks set in to resolve the increasing role of the off-diagonal elements, see Appendix~\ref{app:meaning}.
By construction, the \ac{SOC} operator $\Vop$ couples states with magnetic numbers $M$ and $M\pm1$ owing to $m=0,\pm1$ in Eq.~\eqref{eq:Vop}.
Thus, apart from diagonal $\smmat{0\,0}$, the $\smmat{k\,1}$ multipoles are most important, as indicated by the reduced \ac{EOM}, Eq.~\eqref{eq:EOM_reduced}.
In turn, higher ranks are responsible for the redistribution of populations between states in the high-spin--high-spin block of the density matrix.
If only the total population of the respective manifold is of interest, these ranks can be truncated.
From our simulations, we infer that the more states participate in such kind of dynamics and the more ``equilibrated'' and uniform it is, the easier it can be represented in terms of mean and lower distribution moments, which correspond to lower-rank state multipoles.

We emphasize that the truncation is especially important when myriads of \ac{SF} basis states have to be included without knowing whether they will be essentially populated or not.
The initial prescreening based on the \ac{SF} energies and dipole and \ac{SOC} coupling matrix elements can be employed to lower the computational cost by excluding a notable portion of states prior to propagation; see our previous work.~\cite{Kochetov_JCTC_2022} 
The reason is that one uses correlated many-body states to reduce the need to resolve correlation during propagation.
The analysis of the present study indicates that one can additionally use angular momentum symmetries to reduce the amount of dynamical information and effort.
However, such a reduction becomes prominent only when many high-spin states are involved.
In this case, a single large problem of the $N_\text{spin}\times N_\text{spin}$ size is recast into a smaller $N_\text{SF}\times N_\text{SF}$ problem for every combination of $k$ and $q$.

\section{Conclusions}
\label{sec:conclusions}

In this work, the \ac{SOC}-driven dynamics in different core-excited transition metal complexes have been recast from the state basis to the basis of irreducible spherical tensors.
Further, the \ac{WE} theorem has been used to separate the degrees of freedom into the dynamical part of the interest and the geometrical part that is irrelevant when only total populations of spin states are considered.
Although the direct reduction of information does not lead to closed equations, they can be used for the physically inspired truncation of the maximal ranks and projections in the full \ac{EOM}.

The efficiency of this reduction depends on the coherence degree of the problem and, thus, depends on the system, its initial state, and preparation conditions, i.e., the state prior to interaction with an electromagnetic field and the details of such interaction.
For the highly coherent cases, as the \Ti\ system considered here, no substantial saving in the computational effort can be expected.
When a large number of participating states leads to lower coherence due to quasi-equilibration in the electronic reservoir, this reduction is substantial, and the full dynamics can be rather closely reproduced with a relatively low rank and projection threshold.
For instance, for the \FeWat\ and \CrWat\ systems, the combination of $k_\text{max}=2$ and $q_\text{max}=1$ produces reasonable results -- the limits which are notably lower than the values needed to reproduce the dynamics exactly.
These limits can be rationalized by considering the physics of the problem and the details of the \ac{SOC} operator.
Since the transformation of the \ac{EOM} from the state basis to the basis of spherical tensors replaces a single large problem with a series of smaller problems, the efficiency of the truncated propagation should stand out for a large number of basis states of high multiplicity.

\appendix
\section{Derivation of the particular terms in the \ac{EOM}}
\label{sec:appendix}

The density operator can be transformed from the state basis $\ket{aSM}$ to the basis of purely \ac{SF} functions $\ket{aS}$ for the spatial part and irreducible spherical tensors $\itens{T}{K}{Q}$ for the spin part
\begin{align*}
    &\hat{\rho} = \nonumber\\
    &\sum_{\substack{aS,bS'\\M,M'}} \bra{aS}\hat\pi\ket{bS'}\bra{SM}\hat\sigma\ket{S'M'}\ket{aS}\bra{bS'}\otimes\ket{SM}\bra{S'M'}\nonumber\\
    &= \sum_{aS,bS'}\sum_{K\,Q}\smel{KQ}{aS,bS'}\ket{aS}\bra{bS'}\otimes\itens{T}{K}{Q}(S,S') \, ,
\end{align*}
where
\begin{align*}
    \smel{K\,Q}{aS,bS'} &= \sum_{MM'}\bra{aS}\hat\pi\ket{bS'}\bra{SM}\hat\sigma\ket{S'M'}\times\nonumber\\
    &\times(-1)^{S-M}\sqrt{2K+1}\tj{S}{S'}{K}{M}{-M'}{-Q}\, 
\end{align*}
is the (time-dependent) expansion coefficient -- state multipole.
To derive the \ac{EOM} for these coefficients, we project different terms in \ac{LvN} Eq.~\eqref{eq:LvN} onto the basis operator $\hat1\otimes\itens{T}{k\dagger}{q}(S,S')$ by taking the trace over the spin part $\Tr\{ \ldots\} = \sum_M \ldots$
For instance, for the density operator itself
\begin{align*}
    &\Tr\{(\hat1\otimes\itens{T}{k\dagger}{q}(S,S'))\hat\rho\} =\nonumber\\ 
    &\Tr\big\{\sum_{aS,bS'}\sum_{KQ}\smel{K\,Q}{aS,bS'}\ket{aS}\bra{bS'}\otimes\itens{T}{k\dagger}{q}(S,S')\itens{T}{K}{Q}(S,S')\big\} =\nonumber\\
    &\sum_{aS,bS'}\smel{k\,q}{aS,bS'}\{S\,k\,S'\}\ket{aS}\bra{bS'}\, , 
\end{align*}
where $\{S\,k\,S'\}$ is the triangular delta -- angular momentum coupling selection rule.
To arrive at this result, we have used~\cite{Blum2012}
\begin{equation*}
    \Tr\{\hat1\otimes\itens{T}{k\dagger}{q}(S,S')\itens{T}{K}{Q}(S,S')\}=\delta_{kK}\delta_{qQ}\{S\,k\,S'\}\, .
\end{equation*}
An \ac{SF} element of the projected density matrix then reads
\begin{align*}
    \bra{cS_c}\Tr\{(\hat1\otimes\itens{T}{k\dagger}{q}(S,S'))\hat\rho\}\ket{dS_d}=\nonumber
    \smel{k\,q}{cS_c,dS_d}\{S_c\,k\,S_d\}\, .
\end{align*}
The same expression can be used for its time derivative.

Further,
\begin{align*}
    \hat H_{\rm CI} &= \sum_{\substack{cS_c, dS_d\\M_c,M_d}}\bra{cS_c}\hat H^{\rm SF}_{\rm CI}\ket{dS_d}\braket{S_cM_c}{S_dM_d}\times\\
    &\hspace{1.5cm}\times\ket{cS_c}\bra{dS_d}\otimes\ket{S_cM_c}\bra{S_dM_d}\\
    &=\sum_{cS_c}E^{\rm SF}_c \ket{cS_c}\bra{cS_c}\otimes\hat1\, , 
\end{align*}
where we employed the closure $\sum_{M_c}\ket{S_cM_c}\bra{S_cM_c}=\hat 1$.
Below we omit the $(S,S')$ dependence at spherical tensors for brevity since the considered block of the density matrix uniquely determines it.
Then,
\begin{align*}
    \Tr\{(\hat1\otimes\itens{T}{k\dagger}{q})&\hat H_{\rm CI}\hat \rho\} = \\
    &\sum_{cS_c,eS_e}E^{\rm SF}_c \smel{k\,q}{cS_c,eS_e}\{S_c\,k\,S_e\}\ket{cS_c}\bra{eS_e}
\end{align*}
and the \ac{SF} matrix element of the first term in the commutator on the r.h.s. of Eq.~\eqref{eq:LvN} reads
\begin{equation*}
    \bra{aS}\Tr\{(\hat1\otimes\itens{T}{k\dagger}{q})\hat H_{\rm CI}\hat \rho\}\ket{bS'} = E^{\rm SF}_{{a}} \smel{k\,q}{aS,bS'}\{S\,k\,S'\}\, .
\end{equation*}

Next, for the dipole moment term, fully analogously, we obtain
\begin{gather*}
    \hat{\pmb{\mu}}=\sum_{cS_c,dS_d} \pmb{\mu}^{\rm SF}_{cd}\ket{cS_c}\bra{dS_d}\otimes\hat1\\
    \bra{aS}\Tr\{(\hat1\otimes\itens{T}{k\dagger}{q})\hat{\pmb{\mu}}\hat \rho\}\ket{bS'} = \sum_{cS_c}\pmb{\mu}^{\rm SF}_{ac} \smel{k\,q}{cS_c,bS'}{\{S_c\,k\,S'\}}
\end{gather*}

Finally, for the \ac{SOC} term,
\begin{widetext}
\begin{align*}
    \Vop &= \sum_m (-1)^m \sum_{cS_c,dS_d}\bra{cS_c}\itens{L}{1}{-m}\ket{dS_d}\bra{S_cM_c}\itens{S}{1}{m}\ket{S_dM_d}\ket{cS_c}\bra{dS_d}\otimes \ket{S_cM_c}\bra{S_dM_d} = \\
    &= \sqrt{3}\sum_m (-1)^{S_c-M_c+m}\tj{S_c}{1}{S_d}{-M_c}{m}{M_d}\bra{cS_c}\itens{L}{1}{-m}\ket{dS_d}\redmel{S_c}{\hat S^1}{S_d}\ket{cS_c}\bra{dS_d}\otimes\\
    &\otimes\sum_{KQ}\sum_{M_cM_d}(-1)^{S_c-M_c}\sqrt{2K+1}\tj{S_c}{S_d}{K}{M_c}{-M_d}{-Q}\itens{T}{K}{Q}(S_c,S_d) =\\
    &=\sum_{KQ}\sum_{\substack{cS_c,dS_d\\M_c,M_d}}\sum_m (-1)^{m}\sqrt{3(2K+1)}\tj{S_c}{1}{S_d}{-M_c}{m}{M_d}\tj{S_c}{S_d}{K}{M_c}{-M_d}{-Q}V^m_{cS_c,dS_d}\ket{cS_c}\bra{dS_d}\otimes\itens{T}{K}{Q}(S_c,S_d)\, .
\end{align*}
Here, we have used the definition of the semi-reduced \ac{SOC} matrix element from Eq.~\eqref{eq:SOC_red}.
When inserting it into $\Tr\{(\hat1\otimes\itens{T}{k\dagger}{q})\hat V\hat \rho\}\}$, one obtains a product of three spherical tensors
\begin{align*}
\Tr\{&(\hat1\otimes\itens{T}{k\dagger}{q})\hat V\hat \rho\}\} = \\
&= \Tr\big\{\sum_{KQK'Q'}\sum_{\substack{cS_c,dS_d,eS_e\\M_c,M_d}}\sum_m (-1)^{m}\sqrt{3(2K+1)}\tj{S_c}{1}{S_d}{-M_c}{m}{M_d}\tj{S_c}{S_d}{K}{M_c}{-M_d}{-Q}V^m_{cS_c,dS_d}\smel{K'Q'}{dS_d,eS_e}\ket{cS_c}\bra{eS_e}\otimes\\
&\otimes\itens{T}{k\dagger}{q}(S_c,S_e)\itens{T}{K}{Q}(S_c,S_d)\itens{T}{K'}{Q'}(S_d,S_e)\big\}
\end{align*}
To tackle them, we apply the product rule for spherical tensors~\cite{Ashby_JMC_1994}
\begin{align*}
    \itens{T}{K}{Q}(S_c,S_d)&\itens{T}{K'}{Q'}(S_d,S_e) = \\
    = &\sum_{\mathcal{K}\mathcal{Q}}(-1)^{S_c+S_f+\mathcal{K}-K+K'-\mathcal{Q}}\sqrt{(2K+1)(2K'+1)(2\mathcal{K}+1)}\tj{K}{K'}{\mathcal{K}}{Q}{Q'}{-\mathcal Q}\sj{K}{K'}{\mathcal K}{S_c}{S_e}{S_d}\itens{T}{\mathcal K}{\mathcal Q}(S_c,S_e) \, .
\end{align*}
Thus, $\Tr\{\itens{T}{k\dagger}{q}(S_c,S_e)\itens{T}{\mathcal K}{\mathcal Q}(S_c,S_e)\big\}=\delta_{k\mathcal K}\delta_{q\mathcal Q}\{S_c k S_e\}$.
In addition, we apply the orthogonality rule for the Wigner 3j symbols
\begin{equation*}
    \sum_{M_cM_d}(2K+1)\tj{S_c}{1}{S_d}{-M_c}{m}{M_d}\tj{S_c}{S_d}{K}{M_c}{-M_d}{-Q}=\delta_{K1}\delta_{Qm}\{S_c\,S_d\,K\}
\end{equation*}
As a result, one gets
\begin{align*}
    \bra{aS}\Tr\{&\hat1\otimes\itens{T}{k\dagger}{q}\hat V\hat \rho\}\}\ket{bS'}=\\
    &\sum_m \sum_{dS_d} \sum_{K'Q'} V^m_{aS,dS_d}\smel{K'\,Q'}{dS_d,bS'}(-1)^{S+S'-Q'}\sqrt{3(2k+1)(2K'+1)}\tj{K'}{1}{k}{Q'}{m}{-q}\sj{1}{K'}{k}{S}{S'}{S_d}{\{S\,k\,S'\}}{\{S\,S_d\,1\}}
\end{align*}
\end{widetext}

In the derivation above, we also used permutational and time-reversal symmetries of 3j symbols and selection rules for the projections in these symbols, e.g., $m+Q'=q$. In addition, we utilized the fact that $k$, $K'$, and $S_c+S_d$ are integer numbers which allowed us to simplify the phase factor.

The same procedure can be followed for each operator for the second term in the respective commutators.
We omit the derivation and refer to the result in Eqs.~\eqref{eq:LvN_spherical}-\eqref{eq:ypsilon2}.

\section{Meaning of different $k$ and $q$}
\label{app:meaning}
A physical meaning can be ascribed to state multipoles with different ranks and components;~\cite{Blum2012} we give the most important examples for the discussion in the main text.
An explicit expression for the state multipoles can be obtained from Eqs.~\eqref{eq:tens_exp}~and~\eqref{eq:t_kq}:
\begin{equation}\label{eq:state_multipole}
\begin{split}
    \smel{k\,q}{aS,bS'} &= \sum_{MM'}\rho_{aSM,bS'M'} \times\\ (-1&)^{S-M'+q+2k} \sqrt{2k+1} \tj{S'}{k}{S}{-M'}{-q}{M}\,.
\end{split}\end{equation}

First, for the rank $k=0$ for subblocks with $S=S'$, Eq.~\eqref{eq:state_multipole} is simplified as follows
\begin{equation}\label{eq:diagK0}
    \smel{0\,0}{aS,bS} = \sum_M \rho_{aSM,bS'M'}
    \frac{1}{\sqrt{2S+1}}\,,
\end{equation}
Thus, the diagonal element $\smel{0\,0}{aS,aS}$ describes the population in the spin-free state $\ket{aS}$ scaled with the factor $1/\sqrt{2S+1}$, while off-diagonal elements describe coherence between $\ket{aSM}$ and $\ket{bSM}$.

For $k=1$ and $q=0$ one obtains (note, that only $S>0$ contributes for $k=1$)
\begin{multline}\label{diaK1Q0}
    \smel{1\,0}{aS,bS} =\\
    \sum_{M>0}(\rho_{aSM,bSM}-\rho_{aS(-M),bS(-M)})\frac{M\sqrt{3}}{\sqrt{S(S+1)(2S+1)}}\,.
\end{multline}
Thus, $\smel{1\,0}{aS,aS}$ describes how large is the difference between the populations in the states $\ket{aSM}$ and $\ket{aS(-M)}$.

As all projections $q\neq0$ depend only on the off-diagonal elements of the density matrix in the spin-state basis $\ket{aSM}$, they describe coherences between states, as also discussed in Ref.~\citenum{Blum2012}.
For the rank $k=1$, the projections $q=\pm1$, given by
\begin{multline}\label{eq:diaK1Q1}
    \smel{1\,\pm 1}{aS,bS} =\\
    \pm \sum_{M \lessgtr \pm (S-1)} \rho_{aSM,bS(M \pm 1)} \sqrt{\frac{3(S \mp M)(S \pm M+1)}{2S(S+1)(2S+1)}}\,,
\end{multline}
describe the coherence between $\ket{aSM}$ and $\ket{bS(M\pm1)}$.
Thus, $\smel{1\,\pm 1}{aS,aS}$ describes coherences between $\ket{aSM}$ and $\ket{aS(M \pm 1)}$.
As such direct transitions break time-reversal symmetry, the diagonal values of these multipoles are zero in our case.
Eqs.~\eqref{diaK1Q0} and \eqref{eq:diaK1Q1} can be combined using spin ladder operators $\hat{S}_{\pm}$ and defining $\hat{S}_0=\hat{S}_z$ to obtain
\begin{equation}\label{eq:diagK1ladder}
    \sum_{aS}\smel{1\,Q}{aS,aS} = \sqrt{\frac{3}{S(S+1)(2S+1)}}\Tr{\hat{S}_Q\hat{\rho}}\,, 
\end{equation}
strictly imposing $\sum_{aS}\smel{1\,Q}{aS,aS}=0$.

In the case of $k=2$ with $q=0$ (for simplicity in the following, we set $S=1$ and only consider diagonal elements),
\begin{align}\label{eq:diaK2Q0text}
    \smel{2\,0}{a1,a1} = -\sum_M \rho_{a1M,a1M}
    (2-3M^2)\frac{1}{\sqrt{6}}\,.
\end{align}
A positive value means that the $M=1$ and $M=-1$, on average, have a higher population than the $M=0$ projection, and for a negative value vice versa. 

Moreover, the projections $q=\pm1$ of the rank $k=2$ compare the coherence between $\ket{a10}$ and $\ket{a1(\pm1)}$ to the coherence between $\ket{a1(\mp1)}$ and $\ket{a10}$
\begin{equation}\label{eq:diaK2Q1}
     \smel{2\,\pm 1}{a1,a1} = \frac{1}{\sqrt{2}} (\rho_{a10,a1(\pm 1)}-\rho_{a1(\mp1),a10})\,.
\end{equation}
In turn, $\smel{2\,\pm2}{aS,bS}$ describe the coherence between $\ket{aSM}$ and $\ket{aS(M\pm2)}$
\begin{equation}\label{eq:diaK2Q2}
     \smel{2\,\pm 2}{a1,a1} = \rho_{a1(\mp1),a1(\pm 1)}\,.
\end{equation}

\section*{Data availability statement}
The data that support the findings of this study are available from the corresponding author upon reasonable request.

\bibliography{Spin_dynamics_WE}

\end{document}